\documentclass{article}

\usepackage{arxiv}

\usepackage[utf8]{inputenc} 
\usepackage[T1]{fontenc}    
\usepackage{hyperref}       
\usepackage{url}            
\usepackage{booktabs}       
\usepackage{amsfonts}       
\usepackage{nicefrac}       
\usepackage{microtype}      
\usepackage{lipsum}		
\usepackage{babel}
\usepackage{csquotes}
\usepackage{graphicx}
\usepackage{doi}
\usepackage[x11names]{xcolor} 
\usepackage{graphicx}
\usepackage{subcaption}
 \usepackage{amsmath}

\title{First Linearity and Stability Characterization for CZT Detection System in an e$^+$e$^-$ Collider Environment}

\begin{document}

\maketitle
Leonardo Abbene $^{3}$, Francesco Artibani $^{1,2}$*, Manuele Bettelli $^{5}$, Antonino Buttacavoli $^{3}$, Fabio Principato $^{3}$, Andrea Zappettini $^{5}$, Massimiliano Bazzi $^{1}$, Giacomo Borghi $^{6,7}$, Mario Bragadireanu $^{8}$, Michael Cargnelli $^{4}$, Marco Carminati $^{6,7}$, Alberto Clozza $^{1}$, Francesco Clozza $^{1}$, Luca De Paolis $^{1}$, Raffaele Del Grande $^{1,9}$, Kamil Dulski $^{1}$, Laura Fabbietti $^{9}$, Carlo Fiorini $^{6,7}$, Carlo Guaraldo $^{1}$$^\dagger$, Mihail Iliescu $^{1}$, Masahiko Iwasaki $^{10}$, Aleksander Khreptak $^{11}$, Simone Manti $^{1}$, Johann Marton $^{4}$, Pawel Moskal $^{11,12}$, Fabrizio Napolitano $^{1}$, Szymon Niedźwiecki $^{11,12}$, Hiroaki Ohnishi $^{13}$, Kristian Piscicchia $^{14,1}$, Yuta Sada $^{13}$, Francesco Sgaramella $^{1}$, Diana Laura Sirghi $^{1,8,14}$, Florin Sirghi $^{1,8}$, Magdalena Skurzok $^{11,12}$, Michal Silarski $^{11}$, Antonio Spallone $^{1}$, Kairo Toho $^{13}$, Lorenzo Toscano $^{6,7}$, Marlene Tüchler $^{4}$, Oton Vasquez Doce $^{1}$, Johann Zmeskal $^{4}$$^\dagger$, Catalina Curceanu $^{1}$, Alessandro Scordo $^{1}$

\vspace{1cm}
$^{1}$ \quad Laboratori Nazionali di Frascati, INFN, Via E. Fermi 54, 00044 Frascati, Italy
\\
$^{2}$ \quad Dipartimento di Matematica e Fisica, Università di Roma Tre, Via della Vasca Navale, 84, 00146 Roma, Italy
\\
$^{3}$ \quad Dipartimento di Fisica e Chimica - Emilio Segrè, Università di Palermo, Viale Delle Scienze, Edificio 18, Palermo, 90128, Italy
\\
$^{4}$ \quad Stefan-Meyer-Institut für Subatomare Physik, Dominikanerbastei 16, 1010 Wien, Austria
\\
$^{5}$ \quad Istituto Materiali per l’Elettronica e il Magnetismo, Consiglio Nazionale delle Ricerche, Parco Area delle Scienze 37/A, Parma, 43124, Italy
\\
$^{6}$ \quad Politecnico di Milano, Dipartimento di Elettronica, Informazione e Bioingegneria, Via Giuseppe Ponzio 34, Milano, 20133, Italy
\\
$^{7}$ \quad INFN Sezione di Milano, Via Celoria 16, Milano, 20133, Italy
\\
$^{8}$ \quad Horia Hulubei National Institute of Physics and Nuclear Engineering (IFIN-HH), No. 30, Reactorului Street, Magurele, Ilfov, 077125, Romania
\\
$^{9}$ \quad Physik Department E62, Technische Universität München, James-Franck-Straße 1, Garching, 85748, Germany
\\
$^{10}$ \quad Institute of Physical and Chemical Research, RIKEN, 2-1 Hirosawa, Wako, Saitama, 351-0198, Japan
\\
$^{11}$ \quad Faculty of Physics, Astronomy, and Applied Computer Science, Jagiellonian University, Łojasiewicza 11, Krakow, 30-348, Poland
\\
$^{12}$ \quad Centre for Theranostics, Jagiellonian University , Kopernika 40, Krakow, 31-501, Poland
\\
$^{13}$ \quad Research Center for Accelerator and Radioisotope Science (RARIS), Tohoku University, 1-2-1 Mikamine, Taihaku-kun, Sendai, 982-0826, Japan
\\
$^{14}$ \quad Centro Ricerche Enrico Fermi - Museo Storico della Fisica e Centro Studi e Ricerche “Enrico Fermi”, Via Panisperna 89A, Roma, 00184, Italy
\\
$^{\dagger}$ Deceased. 
\vspace{1cm}

$^*$ Correspondence: francesco.artibani@lnf.infn.it, francesco.artibani@uniroma3.it

\newpage

\begin{center}
\textbf{Abstract}    
\end{center}

The SIDDHARTA-2 collaboration built a new cadmium-zinc-telluride (CZT, CdZnTe)-based X-ray detection system, used for the first time in the DA$\Phi$NE electron-positron collider at INFN-LNF. The aim of this work is to show that these detectors present optimal long- and short-term linearity and stability to perform precise spectroscopic measurements in a collider environment. The spectra used as references for calibration are reported, and the results about the linearity and stability studies are presented. It is also discussed and showed what is the proper function to describe all the effects that alter the Gaussian shape in semiconductors, particularly evident in the CZT case. Good residuals and resolutions were obtained for all the calibrations. In a test run with the source and the collider beam on, it was demonstrated that the calibrations made with beam off are optimal also when the beam is on, and the actual systematics in a physics run were estimated. These promising results show the potentialities of this detector in the high rate environment of a particle collider, and pave the way for the use of CZT detectors in kaonic atoms researches and in accelerators, with applications for particle and nuclear physics.

\begin{center}
\textbf{Dedication}    
\end{center}

\noindent \emph{We dedicate this work to the memory of prof. C. Guaraldo and prof. J. Zmeskal. They have been the initiators of the kaonic atoms measurements campaign at DA$\Phi$NE and J-PARC and pioneered the possibility to use CdZnTe detectors for such measurements. This work would not have been possible without their crucial contribution.}

\section{Introduction}

\noindent Cadmium-Zinc-Telluride is an appealing compound semiconductor to fabricate room temperature X-ray detectors with good energy and timing resolution in a wide energy range (\cite{del_sordo_progress_2009}). Being CZT a high-Z material, his efficiency does not fall down fast, as in the case of silicon which can guarantee good performances only until tens of keV. Furthermore, this promising compound semiconductor, thanks to its high bandgap, ensures ideal performances for room temperature X-ray and $\gamma$-ray spectroscopy at room temperature, without the needing of a cooling system. These unique features, together with the recent improvements on the growth of the crystals to avoid impurities, made CZT one of the most promising materials in the field of semiconductors for X-ray spectroscopy (\cite{abbene_high-rate_2015, abbene_development_2017, abbene_room-temperature_2020, vicini_optimization_czt_2023, abbene_potentialities_2023}). The SIDDHARTA-2 (Silicon Drift Detectors for Hadronic Atom Research by Timing Application) collaboration arranged a new CZT detection system with the aim to perform new measurements at the DA$\Phi$NE collider (\cite{abbene_new_2023, scordo_cdznte_2024}), paving the way for the use of this kind of semiconductor in nuclear and sub-nuclear physics research.

\noindent The collaboration, exploiting the unique beam of low-energy kaons produced in the DA$\Phi$NE collider at the Laboratori Nazionali di Frascati (LNF), is strongly involved in kaonic atoms research (\cite{bazzi_first_2011, bazzi_new_2011, curceanu_kaonic_2020, sirghi_kaonic_helium_2022}). 
Kaonic atoms spectroscopy (\cite{curceanu_frontiers_2023, curceanu_modern_2019, artibani_odyssey_2024}) is particularly important in the field of particle and nuclear physics (\cite{bernard_chiral_2008, cieply_pole_2016}), and has also applications on the cascade models (\cite{torigoe_atomic_cascade_1977}) and astrophysics (\cite{ramos_kaon_condensation_neutron_stars_2000}). From the spectroscopy of the kaonic atoms, together with the radiative de-excitation processes, two important observables linked to the strong interaction at very low energy can be measured in the last transitions: a shift ($\epsilon$) with respect to the purely electromagnetic process and an intrinsic width ($\Gamma$) of the level. These two observables represent unique inputs that theories on low-energy strong interactions must use (\cite{bernard_chiral_2008, cieply_pole_2016}).

\noindent The main goal of the SIDDHARTA-2 experiment is the first determination of shift and width due to strong interactions in the transitions to the 1s level of the kaonic deuterium. This measurement will be crucial to understand the isospin-breaking effects in hadron physics with strangeness (\cite{curceanu_modern_2019}). Together with this groundbreaking goal, the collaboration is also planning to measure the strong interaction in kaonic atoms across the whole periodic table. To accomplish this second goal, innovative X-ray detectors using high Z semiconductors to extend the range of the detectable X-rays to hundreds of keV and MeV: the High Purity Germanium (HPGe) detector (\cite{Bosnar_kaonic_lead_2024}) and the Cadmium-Zinc-Telluride detection system, have been built and tested.

\noindent For what concerns kaonic atoms heavier than hydrogen and deuterium, a series of experiments performed between '70s and '80s, produced an extended set of measurements of shift and width from many kaonic atoms across the periodic table (\cite{friedman_density-dependent_1994, batty_strong_1997}), but recent experiments (\cite{bazzi_kaonic_helium_2009}) measured different values for some atoms, proving that some of these results are not reliable and casting doubts on the present knowledge about kaons interaction with multiple nucleons (K-multiN interaction). Furthermore, with modern technologies, a better precision can be reached, and new, stringent limits can be set for the phenomenological models on K-multiN interaction based on these measurements, with implications that go from particle and nuclear physics (\cite{bernard_chiral_2008, cieply_pole_2016}) to astrophysics (\cite{tolos_strangeness_2020}).

\noindent To measure the strong interaction in kaonic atoms systems in the intermediate-mass range (Al, F, C, S), the SIDDHARTA-2 collaboration has developed a new detection system based on the appealing CZT compound semiconductor. The arrangement of this detection system was particularly challenging because this is the first application in a collider, and this experiment can open the way to new use in this environment. Previous tests already showed the suitable performances of the detector in terms of resolution, efficiency, and timing (\cite{abbene_new_2023, scordo_cdznte_2024}). 

\noindent The study carried on in this paper is to properly control the reliability of the detection system before the physics studies. At first, a study about the short-term stability is essential to show that the system's performance do not depend on environmental factors (temperature, humidity, vibrations, and mechanical noise) and that the hardware and the software systems are stable and do not present different behaviours in various conditions. 

\noindent For the same reasons, it is also important to study the stability over a longer period through different calibration runs. The experiment aims to measure shifts of kaonic atoms with great precision, using data collected over weeks and even months to obtain as much statistics as possible. In addition, the presence of a radioactive source during the data collection could alter the results of the physics researches, being the spectral lines of many sources close to the ones of the various kaonic atoms that the experiment is aiming to measure. For these reasons, it was decided to calibrate the detector using specific calibration runs with the machine off, done approximately once every two weeks. The environmental factors can be particularly significant for the electronic components of the hardware and for the detector themselves, and an accurate study of the performances in different weeks is essential to assure the reliability of the calibration procedure. It is also important to monitor the stability of the system because the apparatus is usually turned off before calibrations, and the data selection for calibration differs slightly from that used for physics runs. These factors could lead to a loss of stability during different runs.

\noindent Finally, the run with the source and the beam on is essential for many reasons. This last step is important to demonstrate that the calibration mode described before is ideal for this kind of detector and to estimate the error that it will introduce on the physical quantities that the experiment aims to measure. Furthermore, it is a strong confirmation to the fact that the detection system can be calibrated with the beam off, and then used in a physics run without any distortion introduced by the high rate environment of a particle collider.
\noindent In Section 2, the accelerator, the detectors and the procedure adopted in the different runs are described, while in section 3, a summary of the data analysis strategy and the results obtained are shown.
\section{Materials and Methods}


\subsection{The DA$\Phi$NE Collider}
\noindent The experiment is placed in the DA$\Phi$NE collider (\cite{milardi_present_2009, Milardi:2021khj, Milardi:2024efr}). DA$\Phi$NE is an electron-positron collider at the Laboratori Nazionali di Frascati (LNF) of INFN (Istituto Nazionale di Fisica Nucleare). This accelerator works at a center of mass energy of 1.020 $GeV$, where the cross-section of e$^+$ e$^-$ interaction is dominated by the $\phi$ resonance production. This machine is unique for kaonic atoms studies because the $\phi$ resonance decays in a pair of charged kaons with a branching ratio of 48\%. The kaons are low-energetic and can be easily stopped in a material to perform kaonic atoms spectroscopy.


\noindent For the DA$\Phi$NE collider, the experimental machine-induced background is dominated by the Touschek effect, an effect that comes from the use of dense beams at relatively low energy to obtain as much luminosity as possible (\cite{boscolo_touschek_2007}). The Touschek effect is a source of background due to the off-energy particles arising from the electromagnetic interactions between particles in a single bunch. This effect produces a large number of off-momentum particles. All of the particles with a momentum dramatically different from the nominal one get lost in the focusing magnets before the interaction point and go out from the beam line, causing a strong electromagnetic background, especially in the region where the CZT detector is placed. For this reason, it is important to study the response of the detector after many physics run, and to compare the performances of the detector when the beam is on and when the beam is off.

\subsection{CZT detection system}
In order to prove that the new CZT X-ray detection technology can be used in the field of kaonic atoms spectroscopy and for applications in physics at colliders, the SIDDHARTA-2 collaboration built and tested a CZT detection system that already showed very good performances in terms of resolutions and timing in two previous tests done in DA$\Phi$NE (\cite{abbene_new_2023, scordo_cdznte_2024}). 

\noindent The CZT detection system consists of eight single $13 \, \text{mm} \times 15 \, \text{mm} \times 5 \, \text{mm}$ quasi-hemispherical CZT detectors enclosed in a thin aluminum box with an 0.27$\mu$m thick aluminum window. These detectors were provided by REDLEN technologies (Canada) and will be referred to as "commercial detectors". In a test done in May, four of the old detectors were replaced by three quasi-hemispherical CZT detectors $10 \, \text{mm} \times 10 \, \text{mm} \times 5 \, \text{mm}$ grown at IMEM-CNR in Parma (Italy) that showed very good performances in preliminary studies, tested for the first time in the DA$\Phi$NE collider.
Both the IMEM-CNR and the commercial detectors present very low-noise gold contacts for anode and cathode electrodes and are coupled to Charge Sensitive Preamplifiers (CSPs) and to a complex readout system that exploits Digital Pulse Processing (DPP), based on an on-line pulse detection and an off-line pulse shape and height analysis of the snapshot waveforms, developed by the section of the collaboration operating in the Dipartimento di Fisica e Chimica (DiFC) Emilio Segré at Palermo University (\cite{Abbene_energy_2013, Gerardi_digital_2014}). These electronics components are enclosed in the aforementioned aluminum box and are the same for all the detectors. To stabilize the temperature of the electronic components (not of the crystals), a FRYKA DLK 402 recirculating chiller working at a temperature of 15°C, was put on the lower side of the aluminum box. The front part of the aluminum box was enclosed with a lead shielding to lower down the intense radiative background caused by the particle losses at the last focusing quadrupole near the interaction point at DA$\Phi$NE that cause a huge background for kaonic atoms researches, being the detector as close as possible to the interaction point.

\begin{figure}[ht]
\begin{center}
\includegraphics[width=10.5 cm]{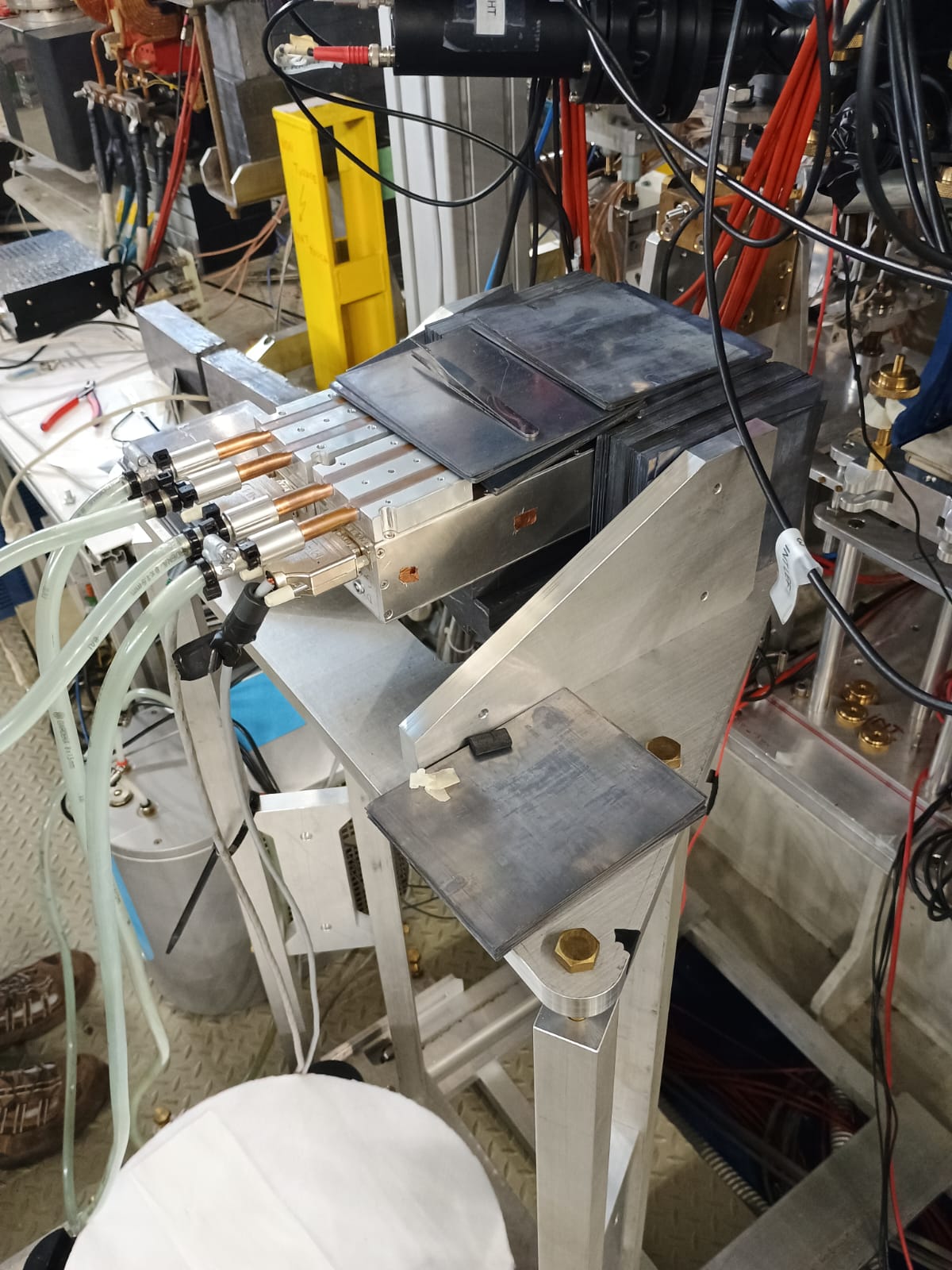}
\caption{Photo of the detection system placed in DA$\Phi$NE interaction region. The detectors in the aluminum box, surrounded by the lead foil shielding, can be seen at the center. On the left-hand side, the chiller system can be found. The detector lays on a special support that allows to change its positions with respect to the beam pipe.\label{det_photo}}
\end{center}
\end{figure}

\noindent A photo of the detection system can be found in Figure \ref{det_photo}.

\noindent The signals coming from the CZT were finally sent to a CAEN DT5780 and V2740 digitizer driven by an original firmware also developed at the DiFC Emilio Segré. 

\noindent Even if a complex trigger system described in (\cite{scordo_cdznte_2024}) was conceived and realized for the experiment, the data acquisitions during the studies presented in this article were performed in self-trigger mode\footnote{All the events that produced a signal on at least one of the detector are saved.}, because in this article only measures concerning calibrations with beam on and with beam off are presented.

\subsection{Calibrations}
\noindent To obtain precise measurements of kaonic atoms observables, it is essential to accurately control the systematics. In addition, the shift due to strong interaction in kaonic atom systems is usually small (except in a few cases), and clearly it cannot be measured if the systematic error is larger than it. For this reason, the study of the stability performances of the detection system, which could contribute dramatically to the systematic error, is extremely important.

\begin{figure}[ht]
\begin{center}
\includegraphics[width=10.5 cm]{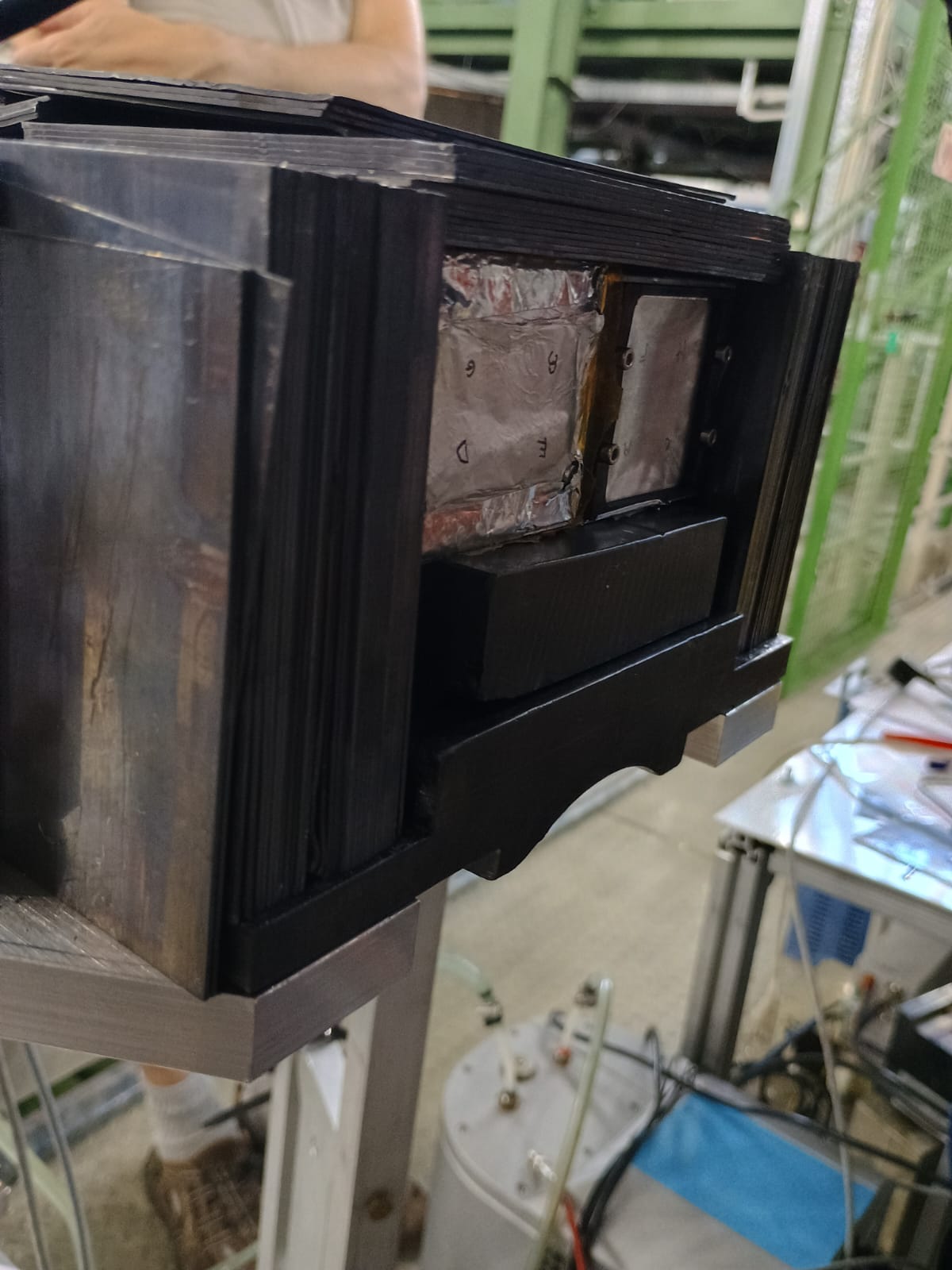}
\caption{Front view of the detection system.\label{front_det_photo}}
\end{center}
\end{figure}

\noindent In this work, the stability was studied exploiting a radioactive source of $^{152}$Eu, placed, with beam off, in front of the detector. A front view of the detector (with a schematic drawing of the source) is reported in Figure \ref{front_det_photo}.

\noindent The chosen radioactive source was the $^{152}$Eu, because it presents two prominent peaks at an energy of ~ 40 keV and ~120 keV that are close to the region of interest of intermediate mass kaonic atoms transition where strong interaction manifests. Following LARA (Library for gamma and Alpha emissions.)\footnote{\url{http://www.lnhb.fr/Laraweb/}} we report in Table \ref{Eu_transitions} the expected energy lines coming from an $^{152}$Eu source in the region inspected by the detector.

\noindent During the data taking of the SIDDHARTA-2 experiment at the DA$\Phi$NE collider, the CZT detector collected data for several months. At first the detector experienced a long phase of optimization of the setup, the HV, and the detector's position. In this study, the calibrations with the same HV on the detector (900V) and the same source ($^{152}$Eu) are reported. A list of the calibrations used in this work can be found in Table \ref{calib_runs_table}.

\begin{table}[ht] 
\begin{center}
\caption{List of all the calibration runs with beam off done between May and July 2024 by the collaboration.\\
In bald are indicated the runs used in this work. \vspace{0.5cm} \label{calib_runs_table}}
\begin{tabular}{cccc}
\toprule
\textbf{Date [DD/MM/YYYY]}    & \textbf{Source}    & \textbf{Detectors} & \textbf{Duration}\\
\midrule
06-07/05/2024    & $^{152}$Eu    & 4 commercial + 3 IMEM-CNR &  25h    \\
09-10/05/2024    & $^{152}$Eu    & 8 commercial &  20h    \\
31/05/2024    & $^{152}$Eu    & 8 commercial &  1h          \\
18-19/06/2024 & $^{152}$Eu & 8 commercial &  16h  \\
\bottomrule
\end{tabular}
\end{center}
\end{table}

\noindent The aim of the study regarded stability in a short-term period (during a single run) and in a long-term period, comparing the results of calibrations done in an interval between a day, twenty days, and a month.

\begin{table}[ht] 
\begin{center}
\caption{Energy transition in the X-ray and $\gamma$-ray regions of $^{152}$Eu. \vspace{0.5cm} \label{Eu_transitions}}
\begin{tabular}{ccccc}
\toprule
\textbf{Energy (keV)}    & \textbf{Intensity (\%)}    & \textbf{Type} & \textbf{Origin} & \textbf{ID}\textsuperscript{1} \\
\midrule
 40.1186 (-)    &  37.7 (5)     &  X K$_{\alpha 1}$ &  Sm           &  Eu1      \\
 121.7817 (3)   &  28.41 (13)   &  $\gamma$         &  $^{152}$Sm   &  Eu3      \\
 344.2785 (12)  &  26.59 (12)   &  $\gamma$         &  $^{152}$Gd   &  Eu5      \\
 39.5229 (-)    &  20.8 (3)     &  X K$_{\alpha 2}$ &  Sm           &  Eu1      \\
 45.4777 (-)    &  11.78 (19)   &  X K$_{\beta 1}$  &  Sm           &  Eu2      \\
 244.6974 (8)   &  7.55 (4)     &  $\gamma$         &  $^{152}$Sm   &  Eu4      \\
 46.6977 (-)    &  3.04 (8)     &  X K$_{\beta 2}$  &  Sm           &  Eu2      \\

\bottomrule
\end{tabular}
\end{center}
\end{table}

\subsection{The in-beam calibration}
\begin{figure}[ht]
\centering
\includegraphics[width=15.5cm]{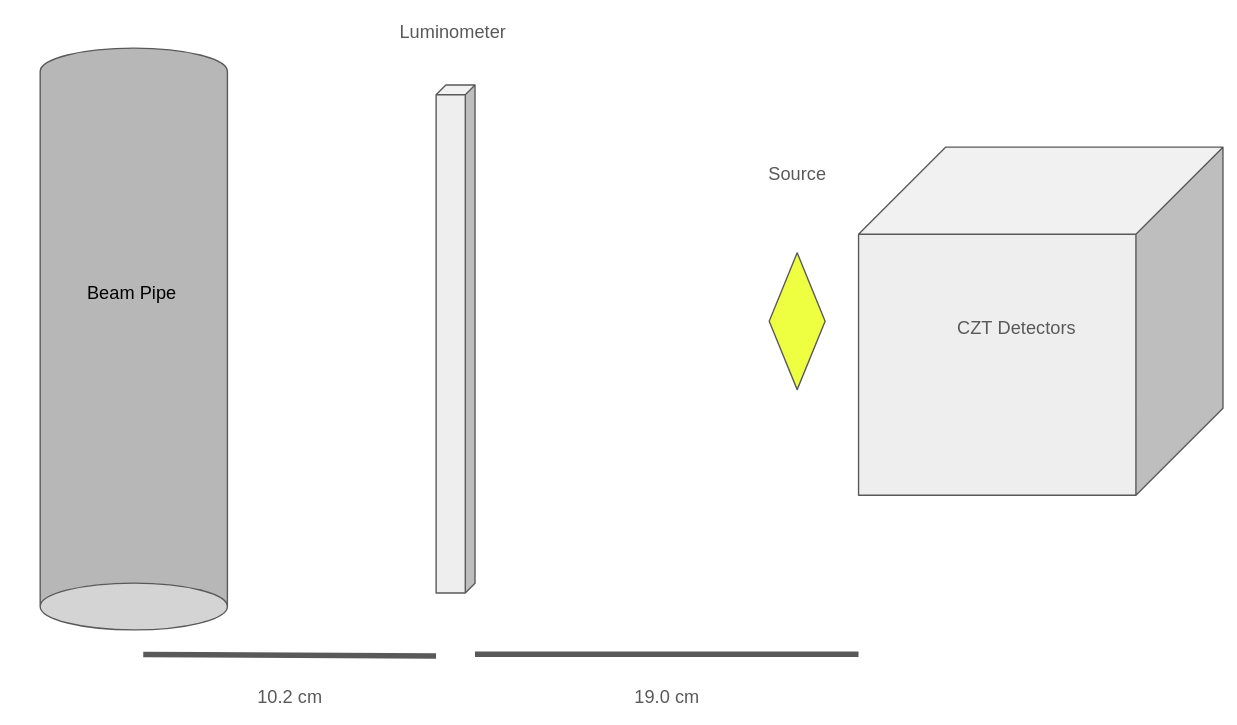}
\caption{Schematic view of the setup during the run with the source.\label{in-beam_scheme}}
\end{figure}

\noindent To demonstrate the validity and reliability of the calibrations, a run with a source and the collider beam on was performed, with the source placed in front of the detection system. The detector was 25 cm far from the Interaction Point (IP) of the DA$\Phi$NE collider, and between them, 10.2 cm from the IP, a plastic scintillator read by two PMTs was placed, working as a luminosity monitor for the whole experiment (\cite{skurzok_characterization_2020, scordo_cdznte_2024}).
A scheme reporting the experimental setup can be seen in Figure \ref{in-beam_scheme}.

\noindent The run lasted ~ 13 hours, and the data were acquired in self-trigger mode.

\noindent This test was particularly useful to confirm that the calibration mode adopted by the experiment is reliable. It was additionally essential to confirm that the system presents no differences when the machine is switched on. Finally, thanks to the study done on this run, it was possible to estimate in a proper way the expected systematic errors confirming the possibility to perform precise spectroscopic measurements of kaonic atoms, crucial in particle and nuclear physics.

\section{Analysis and Results}

\subsection{Peak Fit Function}

\noindent The first discussion that must be carried on is about the peak fit function. Cadmium-Zinc-Telluride detectors at room temperature show a non-negligible tailing effect that must be appropriately described before performing any fit. 

\noindent The two main contributions to the spectral peaks are:
\begin{enumerate}
    \item \textbf{Gaussian function}. The distribution of the energy dispersion caused by the intrinsic resolution of the detector. This effect can be described as a Normal distribution with two parameters: a mean and a dispersion coefficient, the standard deviation.

    \item \textbf{Tail function}. The Gaussian function cannot describe the spectral peaks appropriately if alone. The events that present a wrong recollection of the charge in the semiconductor must also be parameterized together with the good ones.

\noindent In (\cite{gysel_implementation_spectrum_fitting_2003}) an in-depth study on the tailing effect in semiconductor X-ray spectroscopy was carried out, showing that the better form to describe the tail distribution is an exponential. Also in (\cite{borrella_peak_shape_2021}), more specifically on CZT detector, it is claimed that such a function fits well the errors in the event recollection.
\end{enumerate}

\noindent In this work the tail was described using the function suggested in the previous cited articles (\cite{gysel_implementation_spectrum_fitting_2003, borrella_peak_shape_2021}). The fitting function used for a single peak was:
\begin{equation} \label{eq_peak}
    f_{peak}(x) = N \, \times \, \exp\left(-\frac{x-\mu}{2\sigma^2}\right) \, + \, \epsilon \, \times N \, \times \exp\left(\frac{x-\mu}{\beta \sigma}\right) \, \times \, \text{erfc}\left(\frac{x-\mu}{\sqrt{2}\sigma} \, + \, \frac{1}{\sqrt{2}\beta} \right) ,
\end{equation}

where erfc is the complementary error function, used to cut the exponential tail before the Gaussian mean.
In the equation \ref{eq_peak} there are four free parameters: $\mu$ represents the mean of the Gaussian function, i.e. the center of the peak; $\sigma$ represents the standard deviation of the Gaussian function; $\epsilon$ represents the fraction of the height of the tail with respect to the Gaussian; and $\beta$ represents the width of the tail function. The function confirmed to describe in a good way the tailing effect, as visible in the spectrum in Figure \ref{ch7_spec}. 

\subsection{A Typical CZT Spectrum}
\noindent Being the new CZT detection system built by the collaboration the first application that can be found in the literature of this semiconductor in a collider, the response of the detector after being exposed to the high-rate radiation in a collider environment was never explored. In the physics case of kaonic atoms, the control of systematics is essential to determine the effect of strong interaction, as it manifests as a shift with respect to the nominal level calculated with QED and as an intrinsic width of the spectral line. For this reason, an accurate study on the stability of the detectors during the calibration runs was done. 

\begin{figure}[ht]
\centering
\includegraphics[width=17cm]{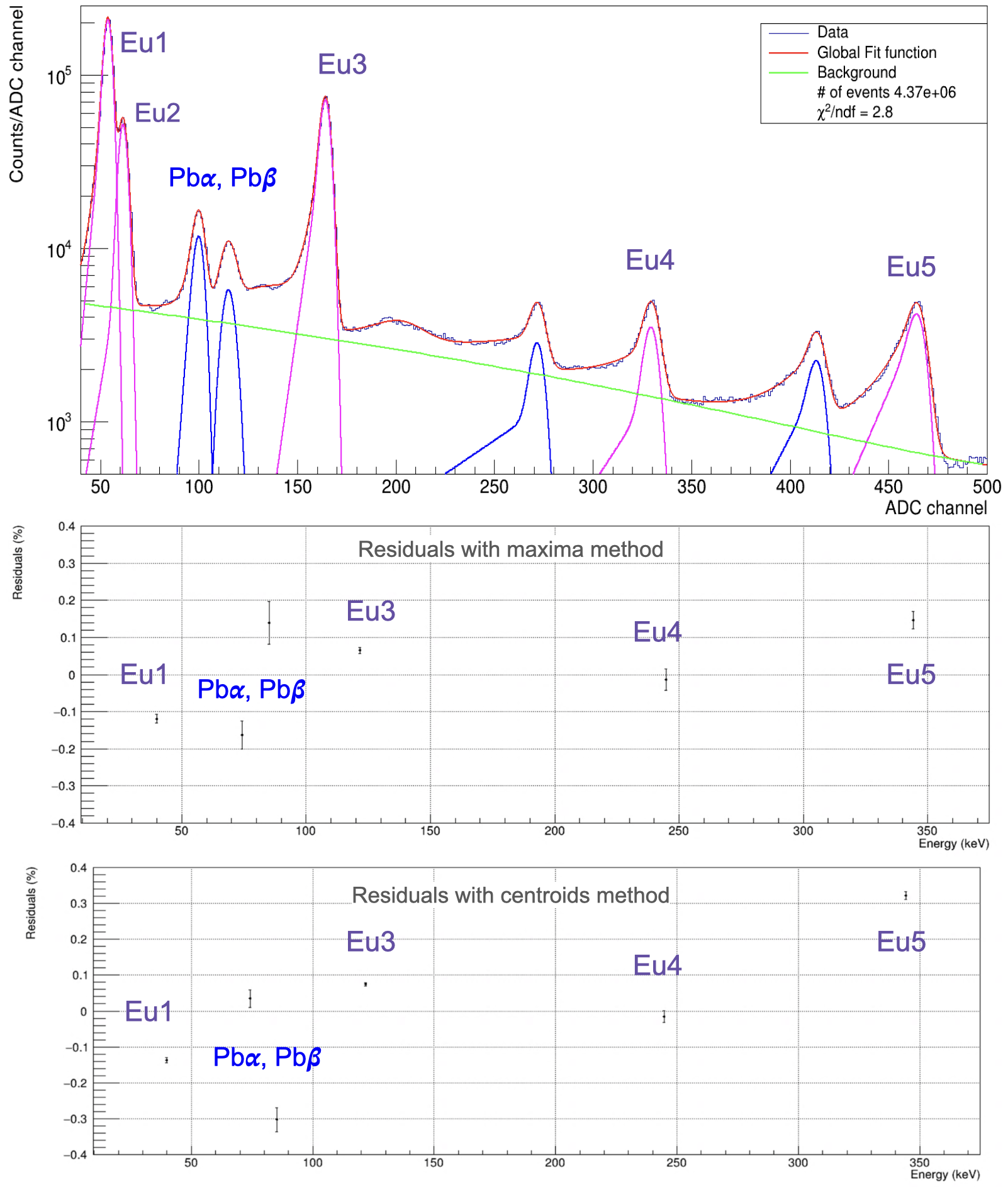}
\caption{Up: Spectrum collected for one of the detectors in the run held from the 5th to the 6th of May 2024.\\Down: Residuals obtained after calibrating the spectrum.\label{ch7_spec}}
\end{figure}

\noindent The calibrations used in this study are presented in table \ref{calib_runs_table}.
A typical spectrum of one of the eight CZT channels is reported in Figure \ref{ch7_spec}, together with the relative residuals 
\begin{equation} \label{formula relative res}
    Residuals(\%) = \frac{(E_{true} - E_{meas})}{E_{true}}.
\end{equation} 

\noindent The residuals are reported for two different methods of calibrations, to emphasize the small differences at high energy.

\noindent In this spectrum, several peaks were identified. Five peaks come from the Europium source (labeled as described in Table \ref{Eu_transitions}), and four peaks come from the interactions between the lead shielding that surrounds the detectors and the $\gamma$-rays from the source. These last four have been carefully studied to avoid systematics in the calibrations due to peaks misidentification. The first two peaks, at ~ 75 keV and ~85 keV, come from the de-excitation of the lead in the shielding after the interactions with high-energy photons. The lead K$_{\alpha1}$, K$_{\alpha2}$, K$_{\beta1}$, K$_{\beta2}$, and K$_{\beta3}$ transitions have indeed an energy of respectively 74.969 keV, 72.805 keV, 84.938 keV, 87.300 keV, and 84.450 keV. The two peaks at higher energies ($\sim$ 200 keV and $\sim$ 300 keV) were identified as backscattering peaks due to the Compton effect. This process happens when high-energy $\gamma$-rays coming from the Europium decay (in particular the transition at ~ 964 keV that is not enough energetic to enable pair production) interact with the shielding through Compton scattering and are then captured in the detector. 

\noindent During a run performed at the end of the data-taking in DA$\Phi$NE as post-calibration, with a different setup without the lead shielding, it was proved that the four peaks coming from interactions of high-energy $\gamma$-rays with the lead disappear.
To perform the calibration, it was decided to use three of the five visible peaks of the Europium, the ones labeled as Eu1, Eu3, and Eu4. The choice was made because the peak Eu2 is not always detectable by all the detectors, given the fact that it is only 5 keV away from the higher Eu1 peak. The peak Eu5, instead, presents strong tailing and distortion effects, but it can be fitted anyway, and it was chosen to be used as a peak to cross-check the calibration.

\noindent Following studies carried on for Cadmium-Telluride (CdTe), another semiconductor with a heavy tail effect (\cite{redus_characterization_cdte_2009}), the reference energy (or ADC) taken from the fit should not be the mean of the Gaussian component but the maximum of the function $f_{peaks}$. Indeed, it was demonstrated that the maxima and the centroids of the Gaussians are reasonably coincident for the peaks with small tailing effect (the ones in the region of interest), but it avoids the possible systematics introduced by the strong tailing effect in these kinds of detectors at higher energies (300 to 500 keV). Also in the case of the CZT detection system, this behaviour was confirmed. As an example, in Figure \ref{ch7_spec}, the resulting residuals when calibrating on of the detectors using centroids and maxima methods are presented. In the figure, it is clear that the residuals for the europium peaks below 300 keV are almost coincident, while in the case of the peak at 344 keV, at higher energies, and so with a higher tailing effect, the residuals using the centroid method is double with respect to the one taking the maxima method. This behaviour was common to all the detectors. 

\noindent In a similar way, clearly, the resulting resolution is not the one of the Gaussian, but the FWHM of $f_{peaks}$, which must also take into account the tailing effect.

\subsection{Short Term Stability}

\begin{figure}[ht]

    \centering
    \begin{minipage}{0.45\textwidth}
        \centering
        \includegraphics[width=\textwidth]{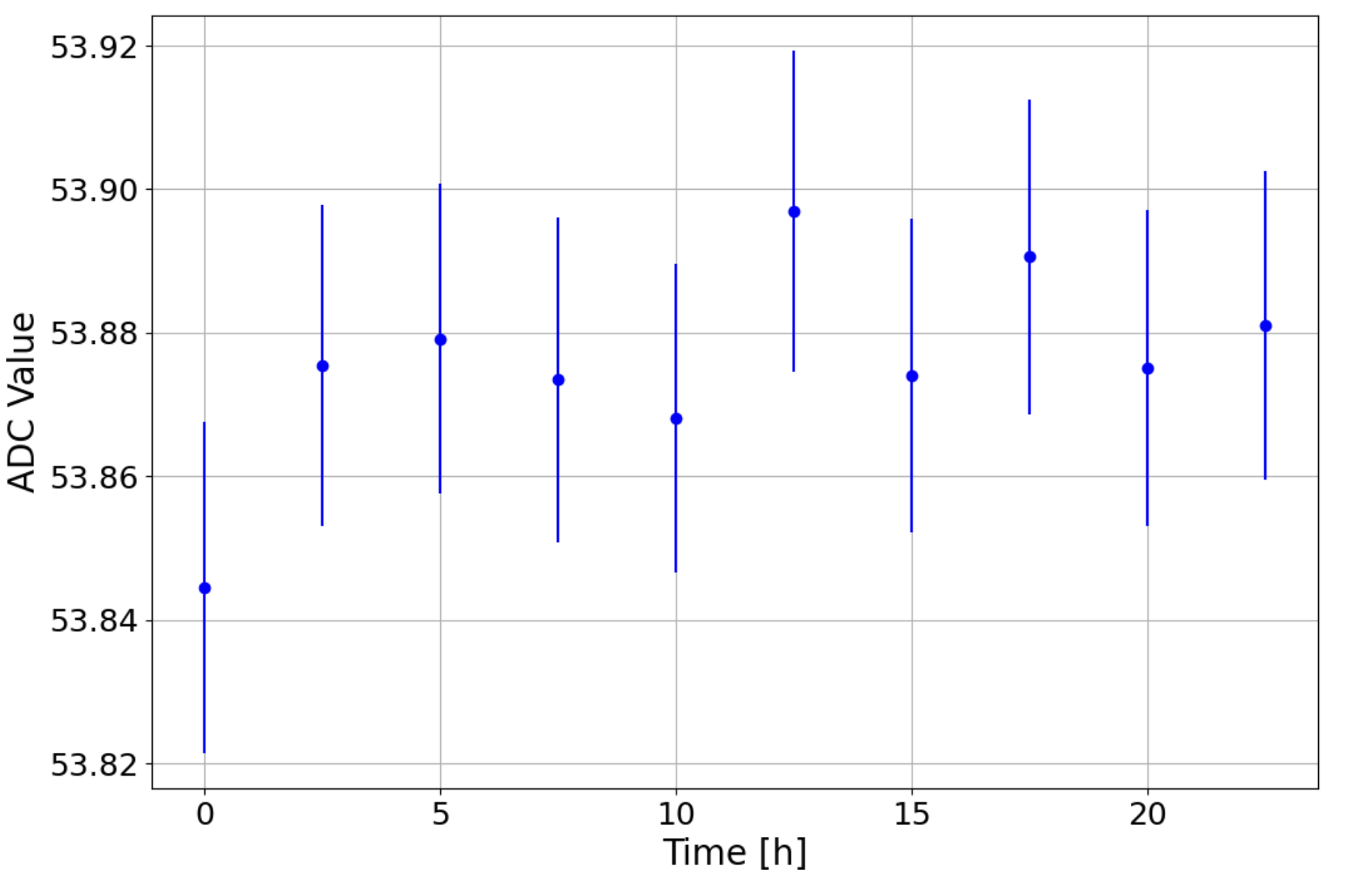}
    \end{minipage}
    \hfill
    \begin{minipage}{0.45\textwidth}
        \centering
        \includegraphics[width=\textwidth]{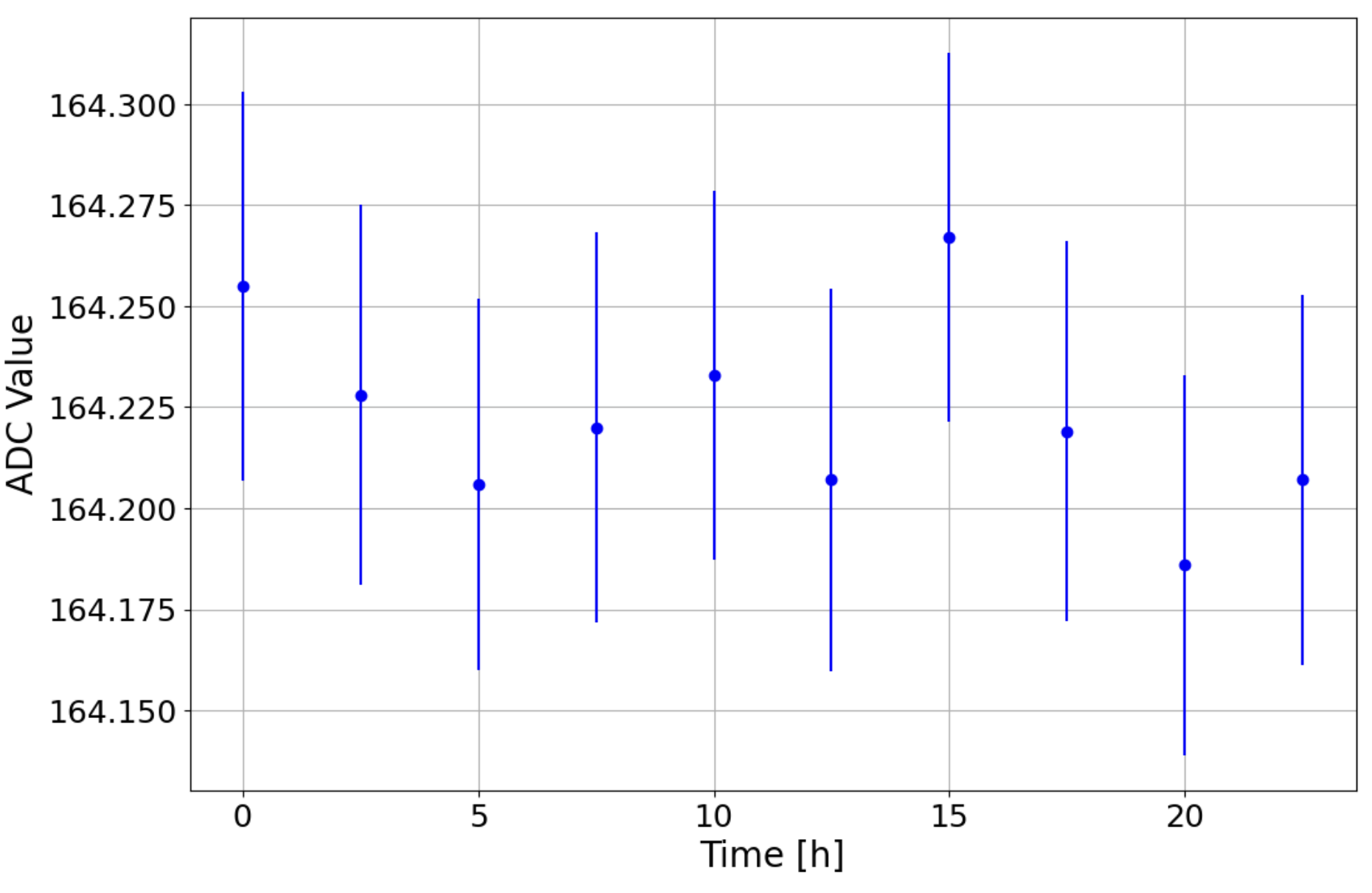}
    \end{minipage}
    \vfill
    \begin{minipage}{0.45\textwidth}
        \centering
        \includegraphics[width=\textwidth]{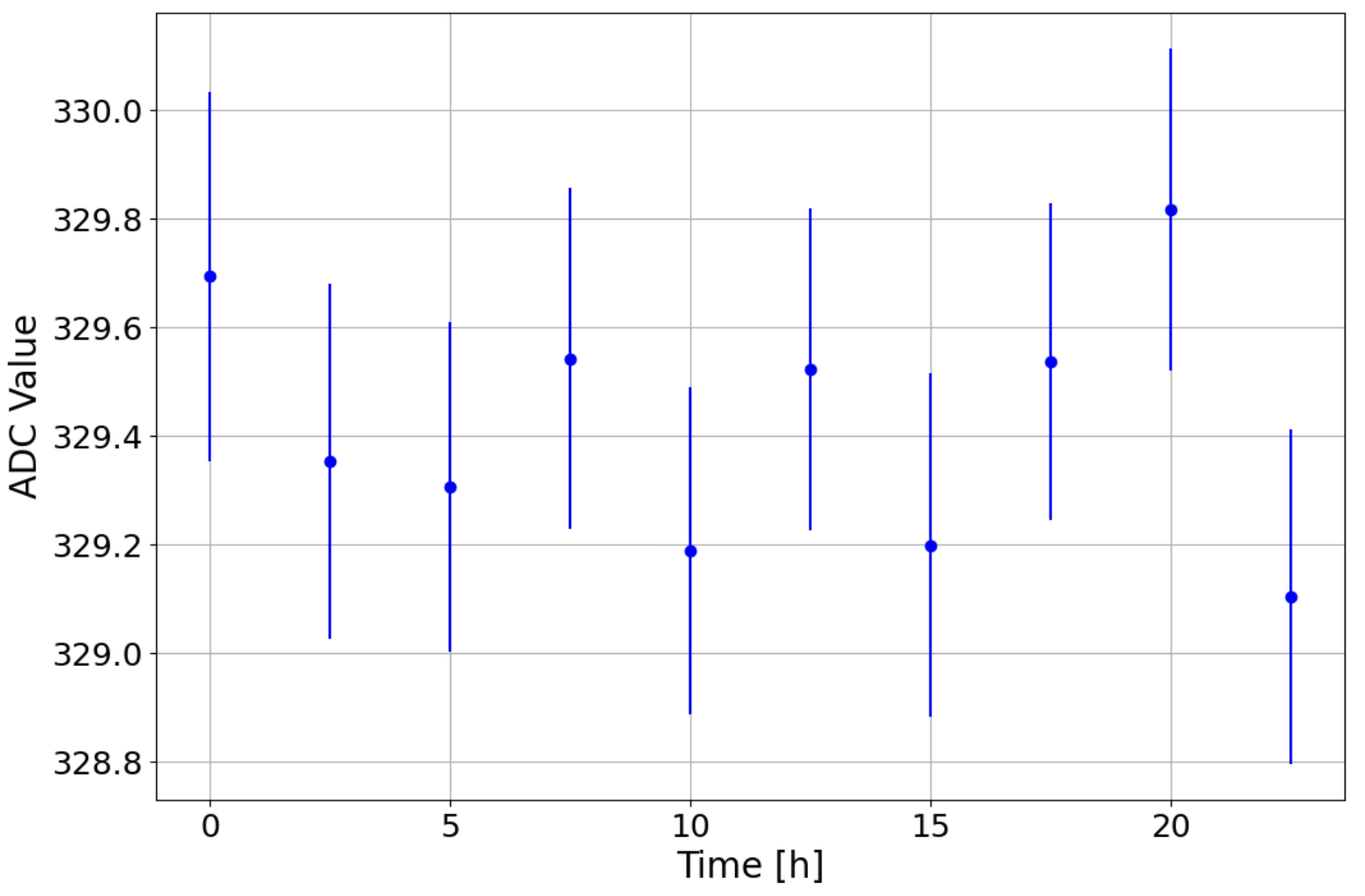}
    \end{minipage}
    \hfill
    \begin{minipage}{0.45\textwidth}
        \centering
        \includegraphics[width=\textwidth]{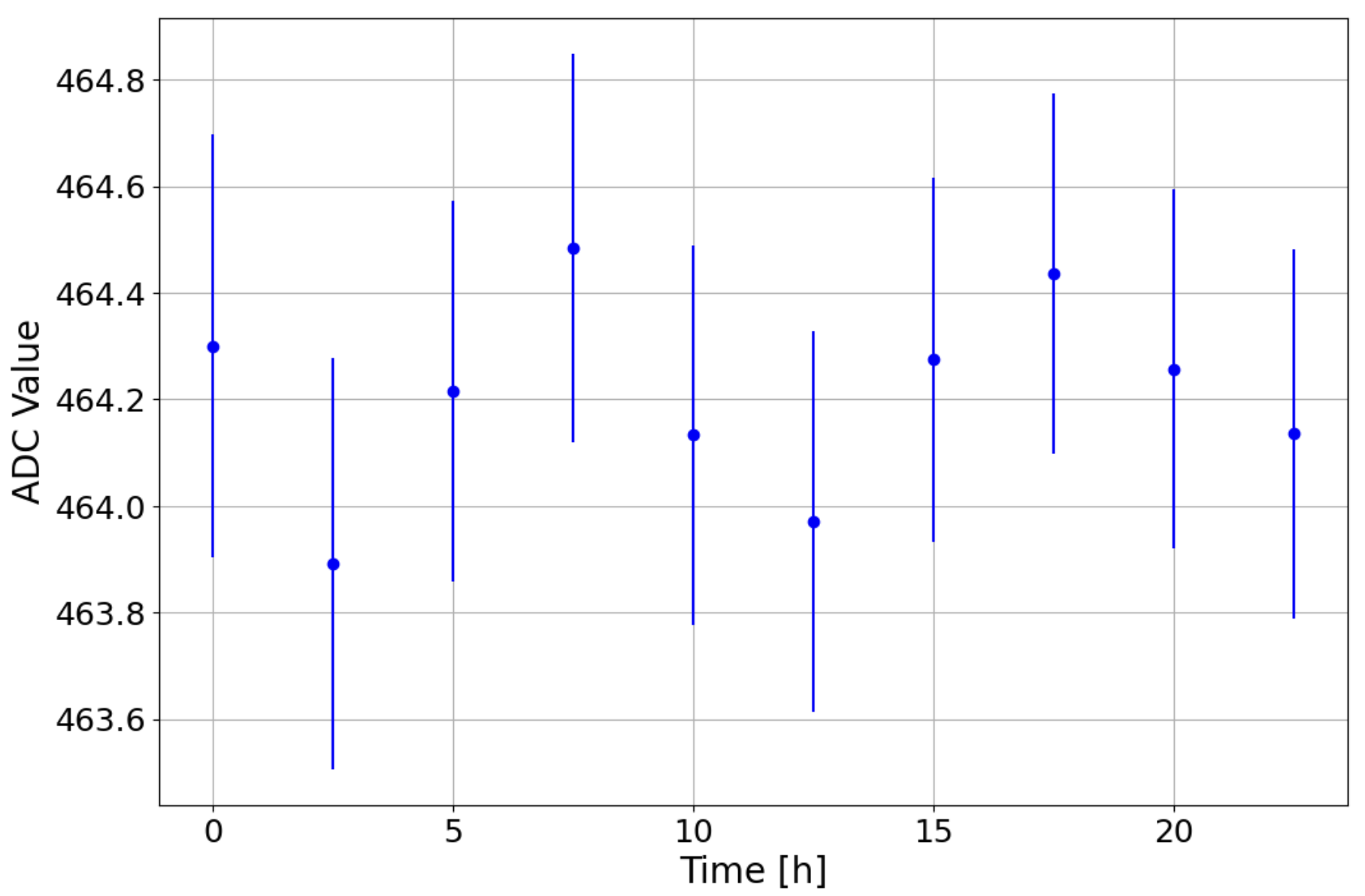}
    \end{minipage}
    \caption{Values of the ADC channel of the $^{152}$Eu peaks obtained for one of the detectors in the different datasets for the peak at 40 keV (up left), at 121 keV (up right), at 244 keV (down left), and at 344 keV (down right).}
    \label{fig:ADC values per hour}
\end{figure}

\noindent The stability during a single calibration run was studied by dividing the longest calibration run, held between 6th and 7th of May 2024, in ten equal small datasets of 2.5 hours each. Every small dataset was fitted with the procedure previously described. In Figure \ref{fig:ADC values per hour}, as a reference, the ADC values of the peaks for one of the detectors are reported. 

\noindent The plot shows that the detector is extremely stable during a 24-hour run, showing a <‰ dispersion between the 2.5-hour dataset. This study confirmed the good stability of the detector and the electronics.

\subsection{Long Term Stability}

\begin{figure}[ht]

    \centering
    \begin{minipage}{0.45\textwidth}
        \centering
        \includegraphics[width=\textwidth]{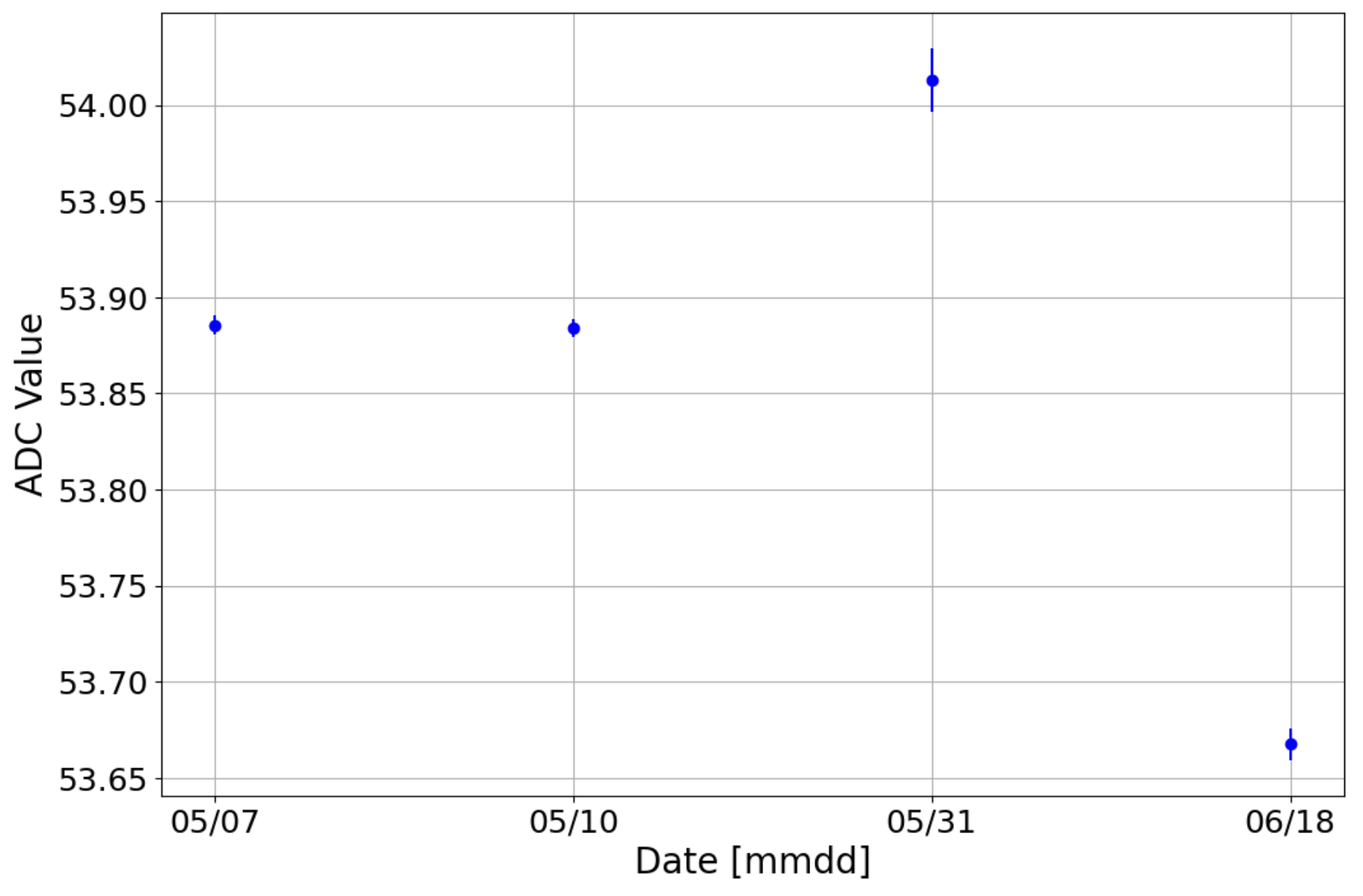}
    \end{minipage}
    \hfill
    \begin{minipage}{0.45\textwidth}
        \centering
        \includegraphics[width=\textwidth]{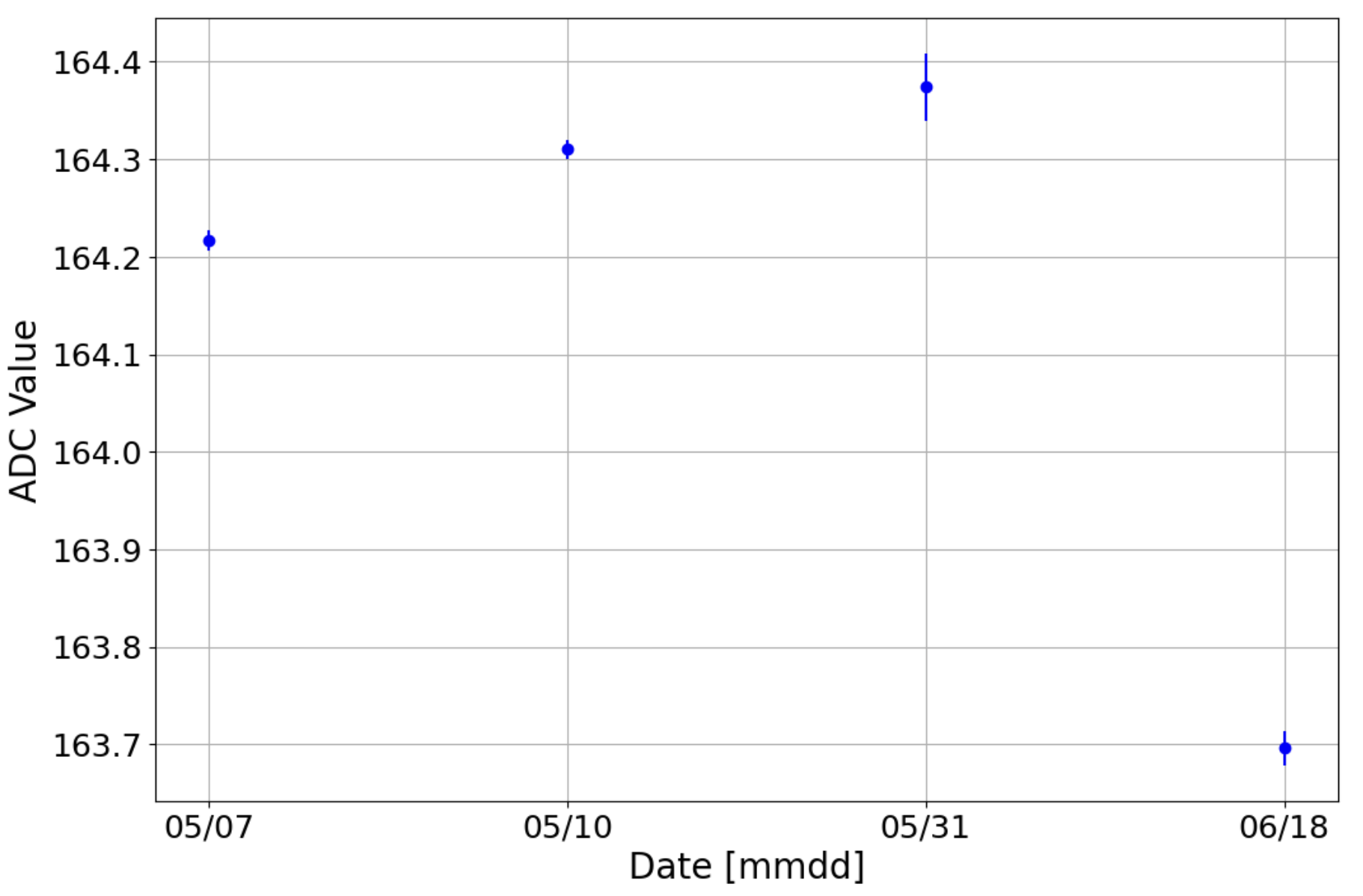}
    \end{minipage}
    \vfill
    \begin{minipage}{0.45\textwidth}
        \centering
        \includegraphics[width=\textwidth]{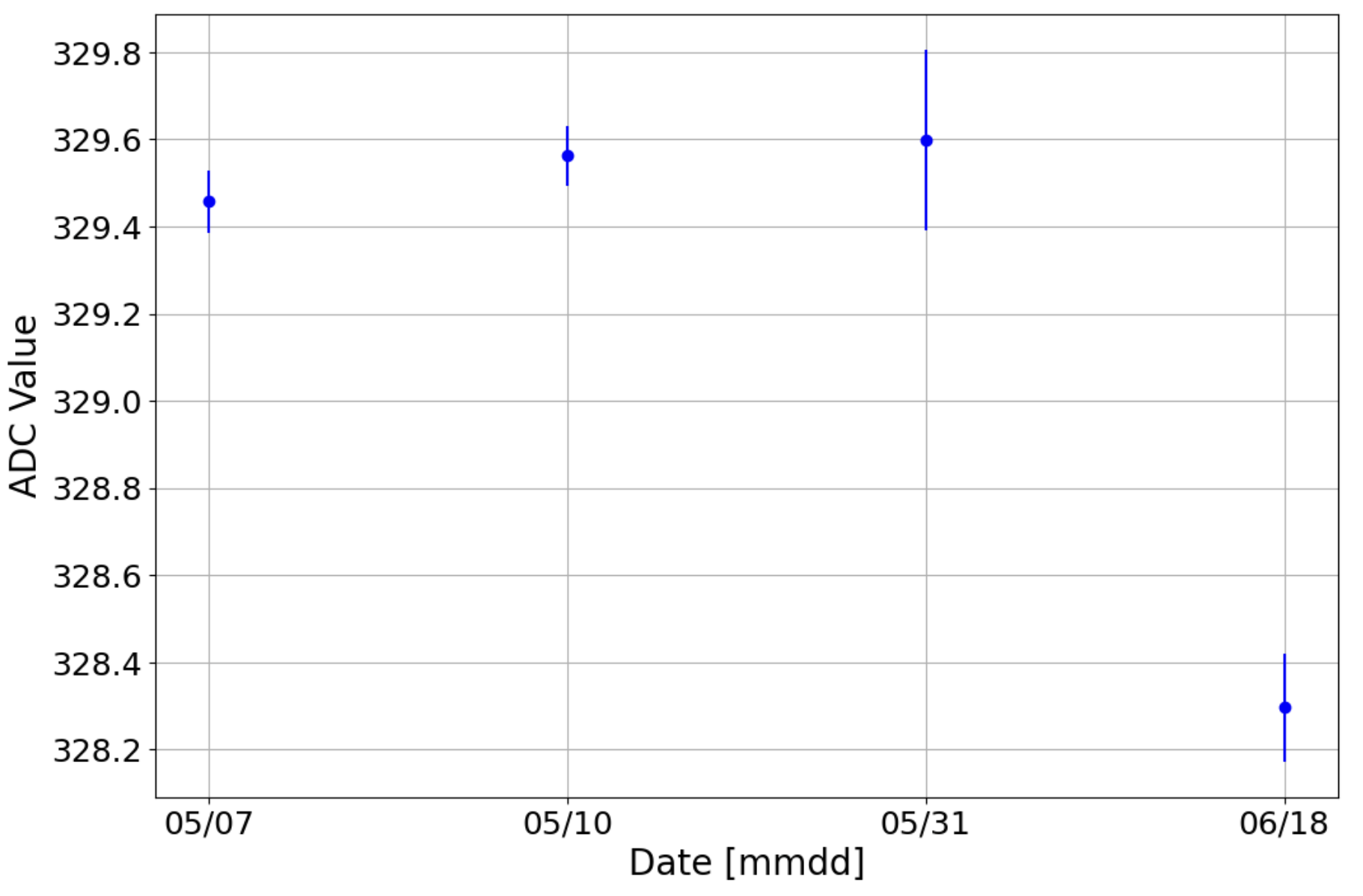}
    \end{minipage}
    \hfill
    \begin{minipage}{0.45\textwidth}
        \centering
        \includegraphics[width=\textwidth]{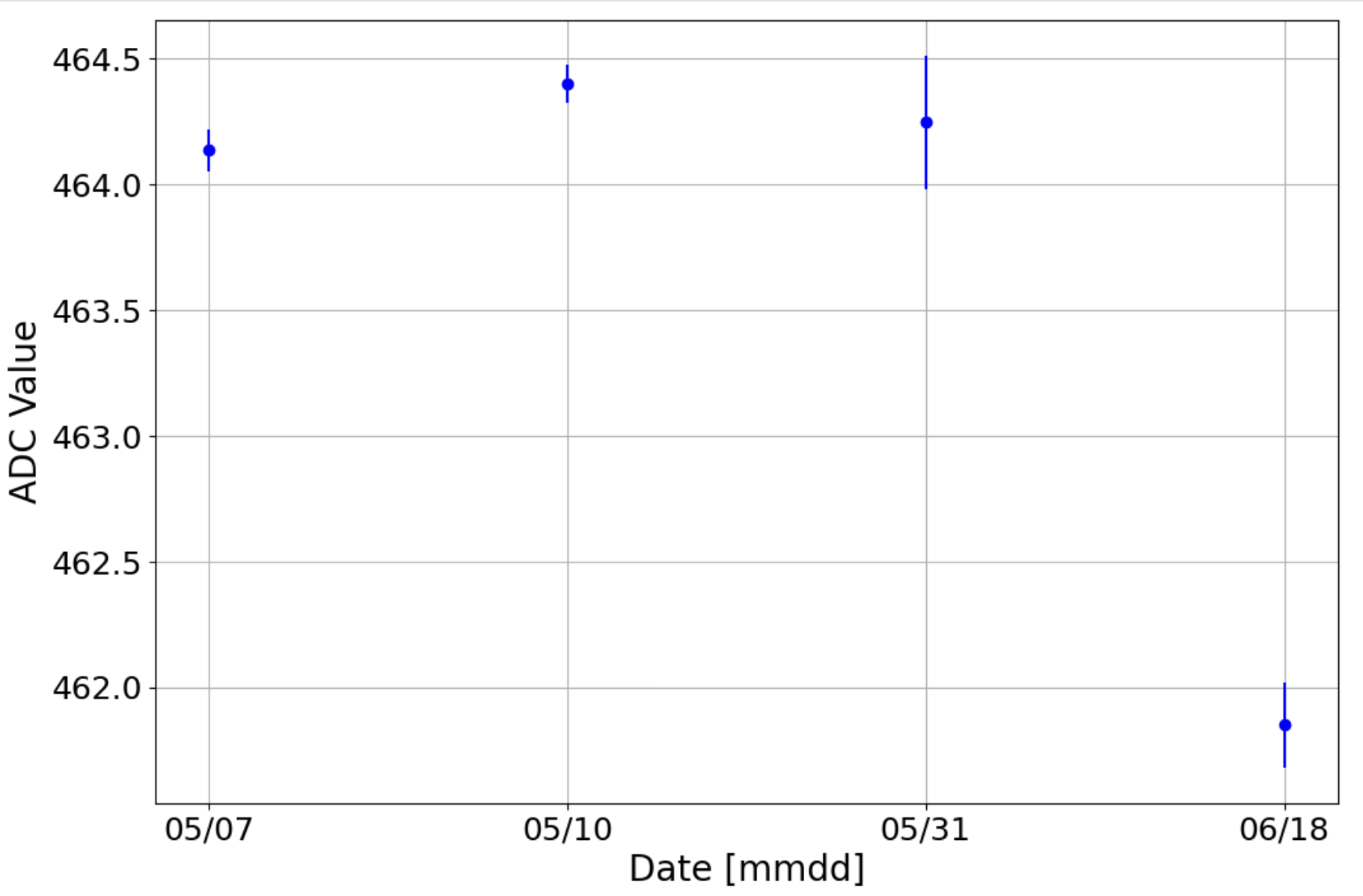}
    \end{minipage}
    \caption{Position of the ADC Maxima for the $^{152}$Eu peaks at 40 keV (up left), 121 keV (up right), 244 keV (down left), and 344 keV (down right) for different dates after the calibrations of one of the detectors.}
    \label{fig:ADC values per date}
\end{figure}

\noindent The long-term stability was carried out studying the dataset in Table \ref{calib_runs_table}. In Figure \ref{fig:ADC values per date}, the ADC values over the taken calibrations are shown. The ADC values of the Europium peaks change for less than 1\%, confirming the good stability of this system.

\begin{figure}[ht]

    \centering
    \begin{minipage}{0.45\textwidth}
        \centering
        \includegraphics[width=\textwidth]{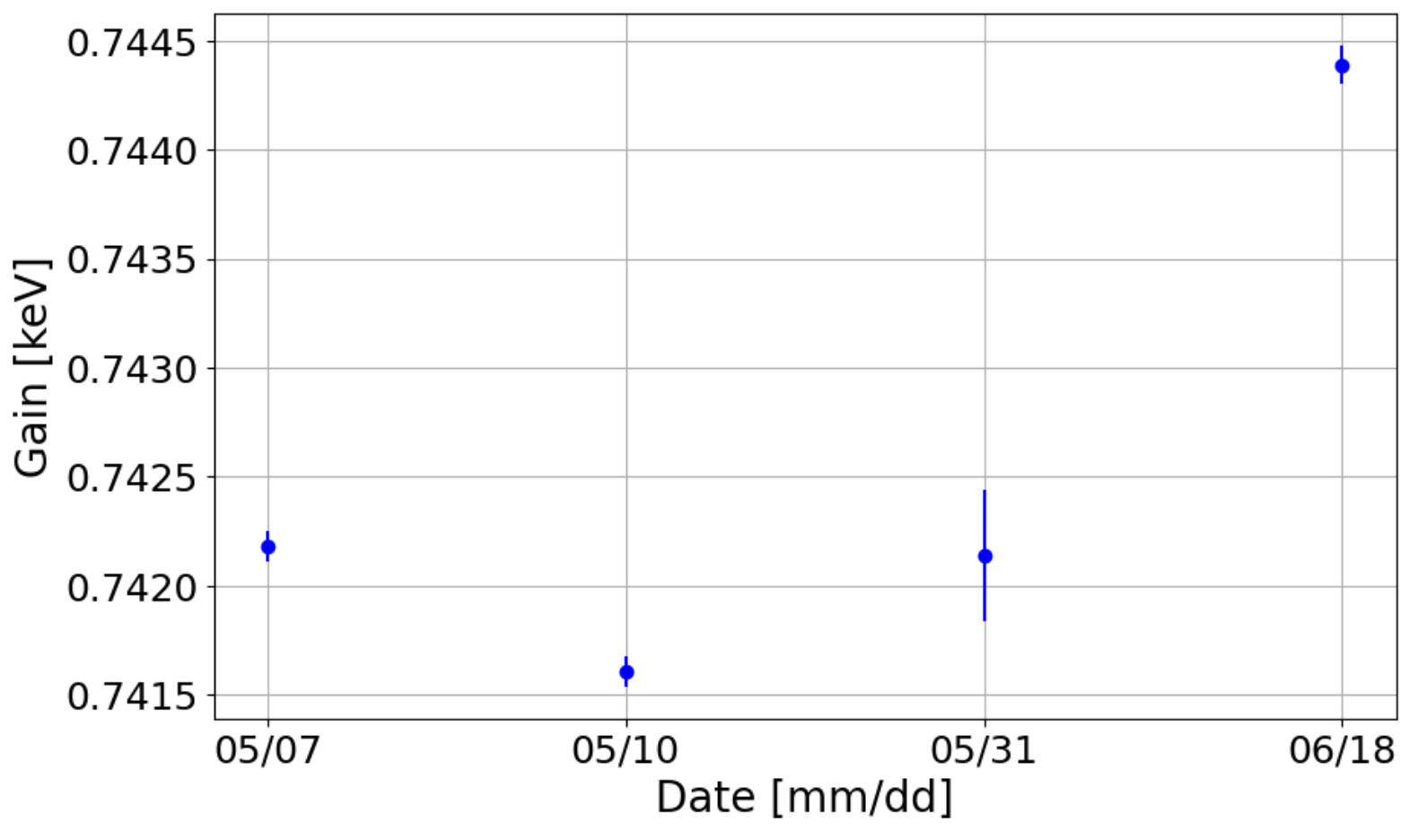}
    \end{minipage}
    \hfill
    \begin{minipage}{0.45\textwidth}
        \centering
        \includegraphics[width=\textwidth]{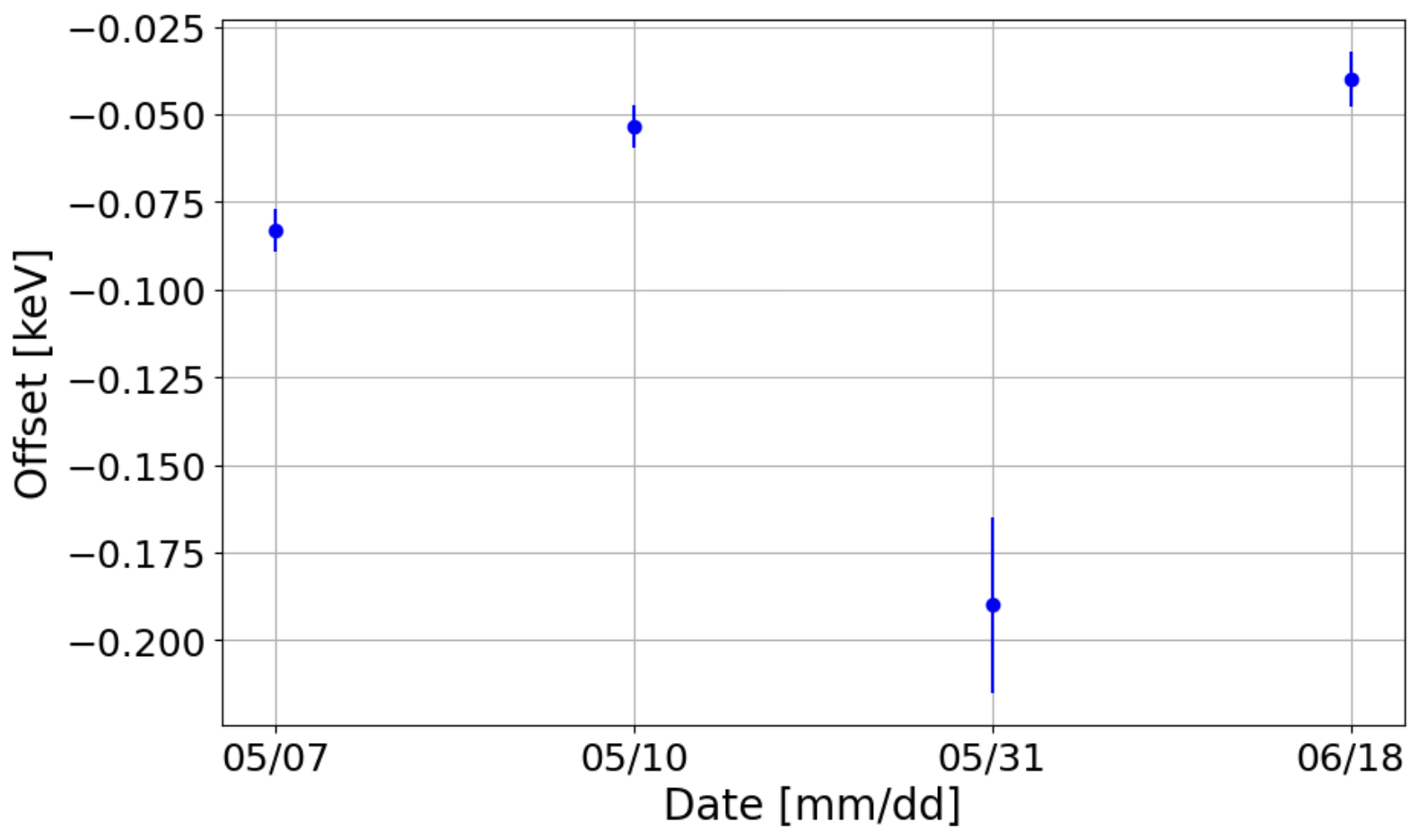}
    \end{minipage}
    \caption{Gains and offsets values for the different calibration runs.}
    \label{fig:gains and off per date}
\end{figure}

\noindent Looking at the gains and offsets obtained after the fit for the same channel, reported in Figure \ref{fig:gains and off per date}, the slight decreasing of the ADC values between the run of 31st of May 2024 and the one of 18th of June 2024 results in a small increasing of the gain (3‰), confirming the reliability of the calibration system in detecting even small fluctuations during data taking.

\begin{figure}[ht]

    \centering
    \begin{minipage}{0.45\textwidth}
        \centering
        \includegraphics[width=\textwidth]{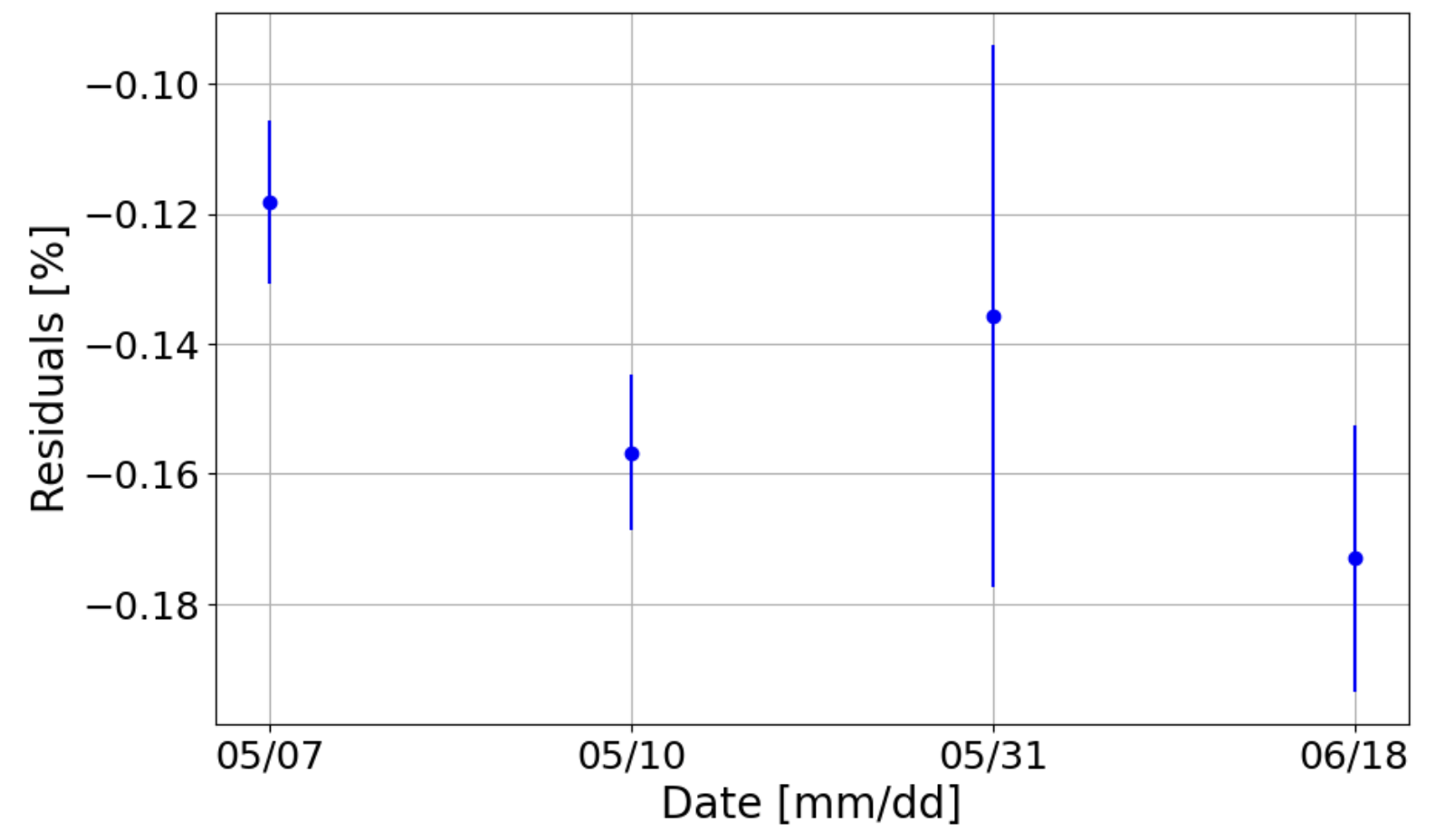}
    \end{minipage}
    \hfill
    \begin{minipage}{0.45\textwidth}
        \centering
        \includegraphics[width=\textwidth]{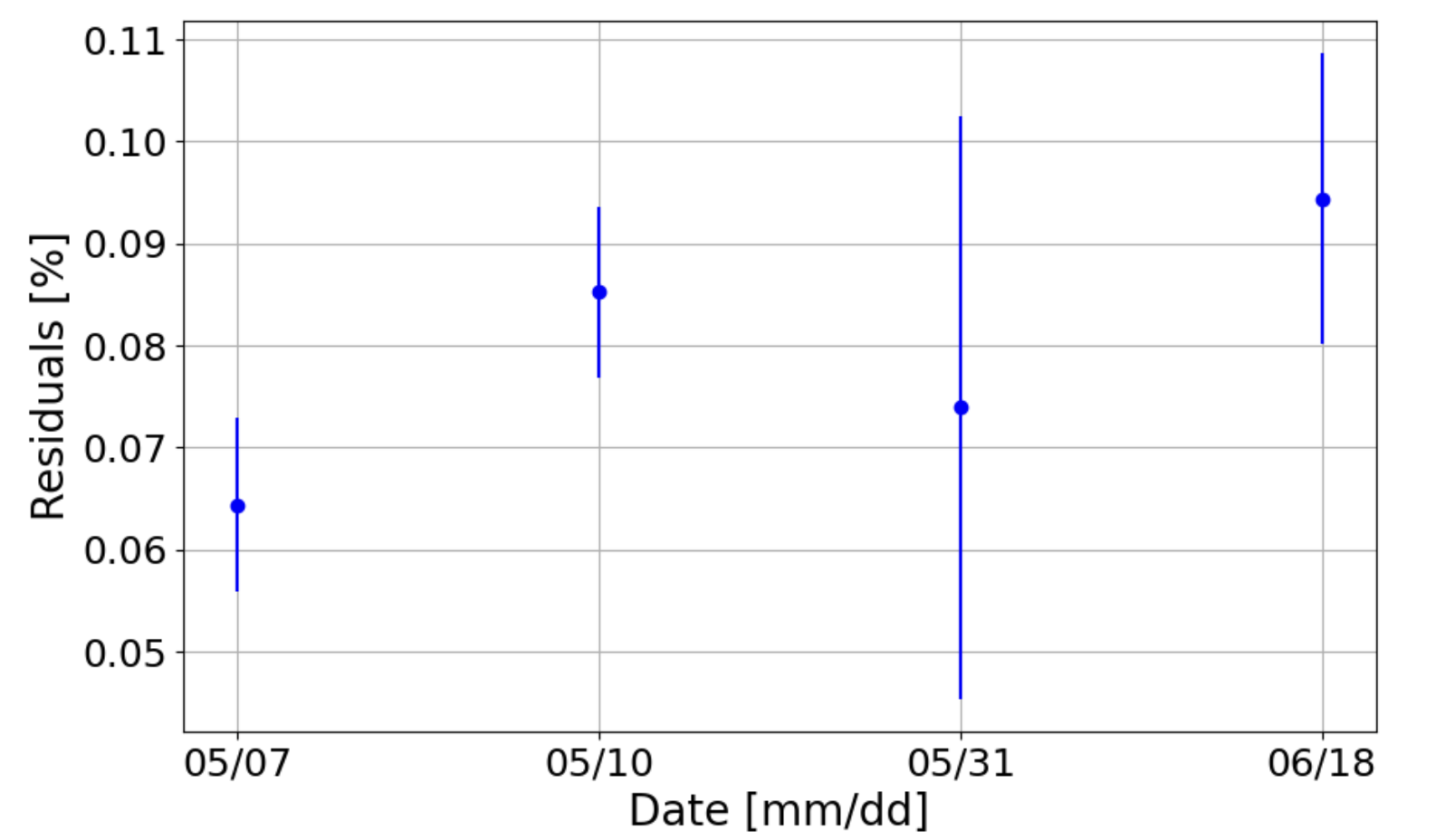}
    \end{minipage}
    \vfill
    \begin{minipage}{0.45\textwidth}
        \centering
        \includegraphics[width=\textwidth]{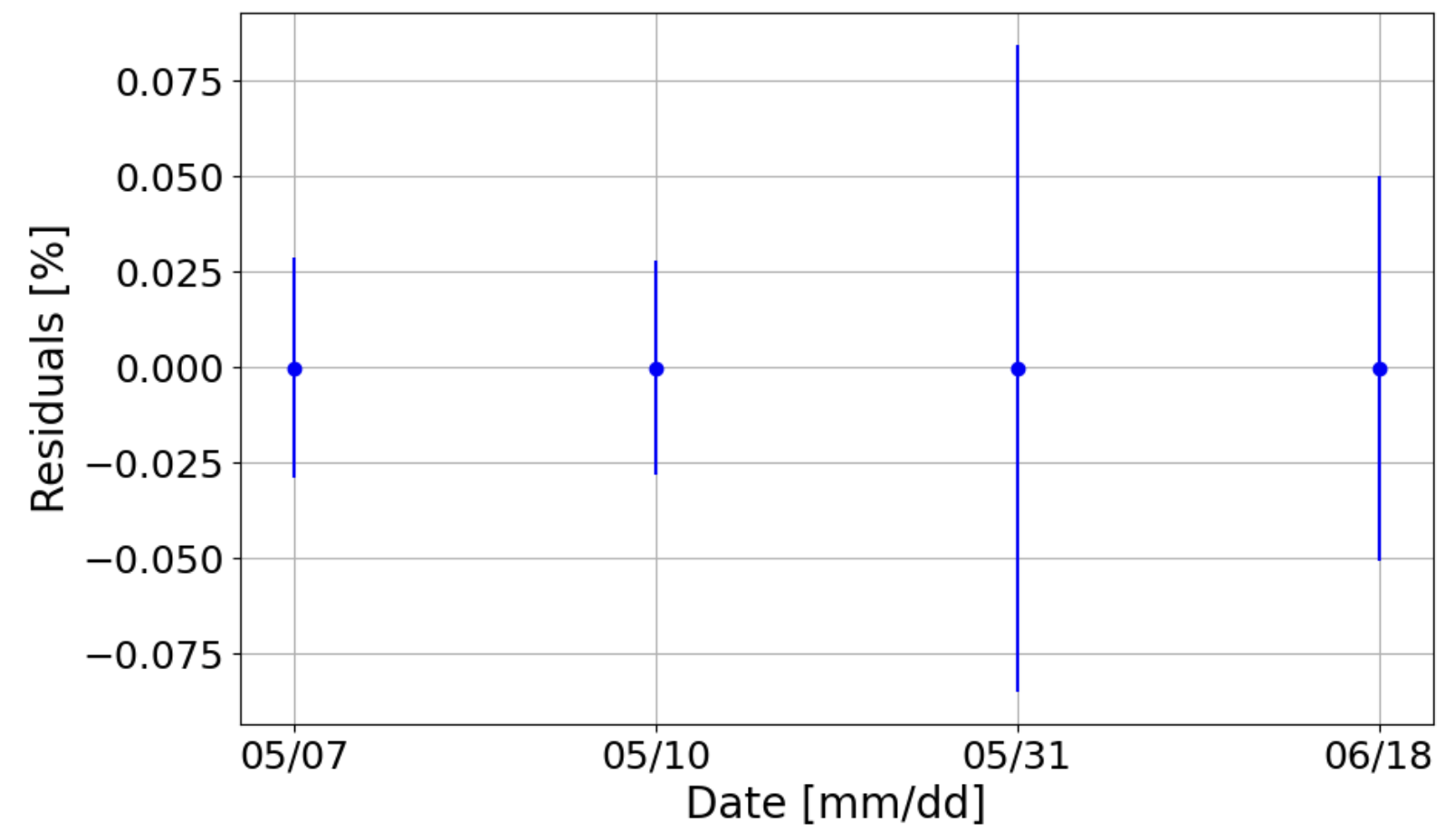}
    \end{minipage}
    \hfill
    \begin{minipage}{0.45\textwidth}
        \centering
        \includegraphics[width=\textwidth]{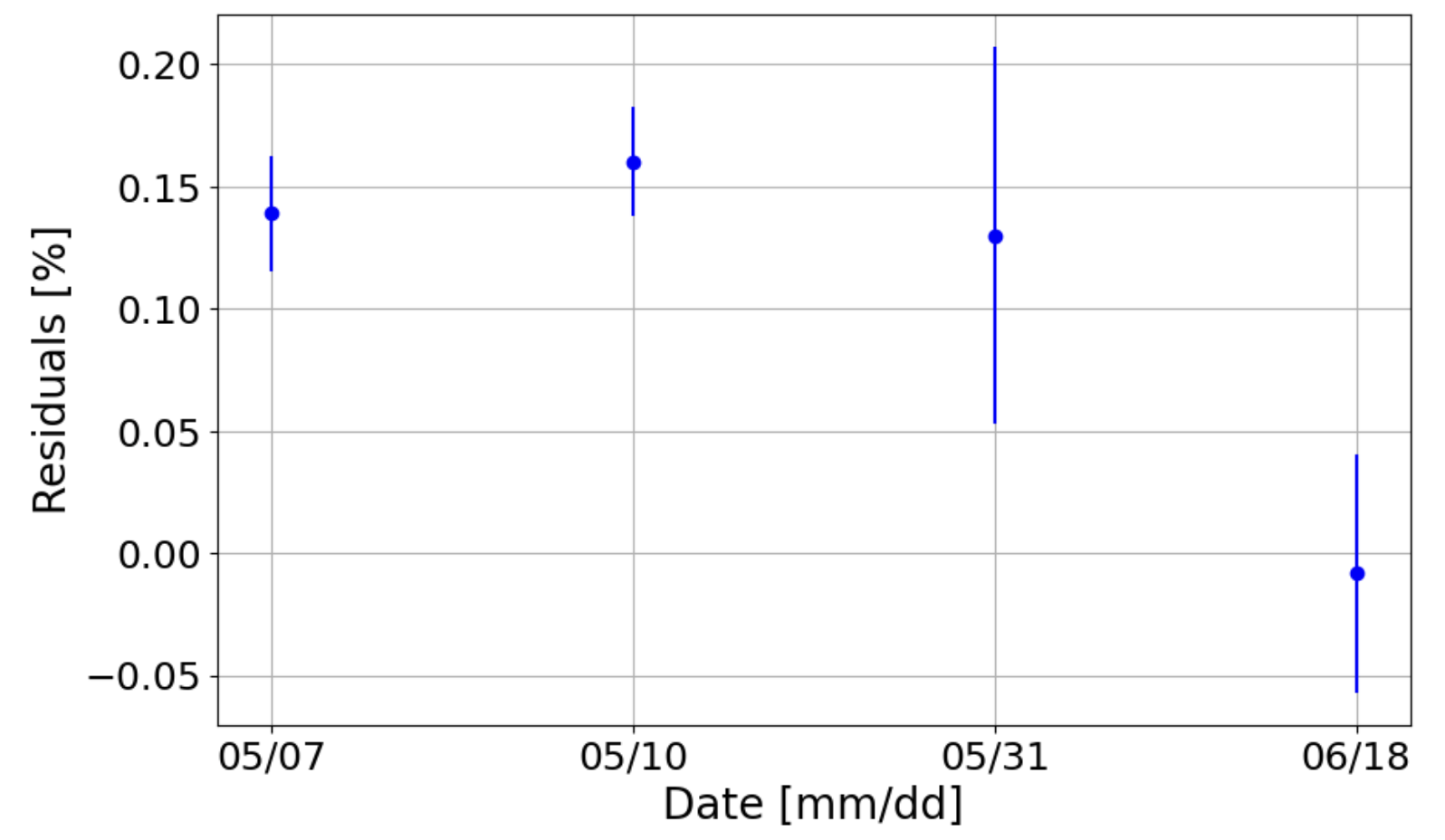}
    \end{minipage}
    \caption{Residuals of the peaks for one of the detectors obtained after the calibrations in different dates for the peak at 40 keV (up left), at 121 keV (up right), at 244 keV (down left), and at 344 keV (down right).}
    \label{fig:Residuals values per date}
\end{figure}

\noindent Putting all together, the detectors showed very good performances: all the residuals after the calibrations were shown to be lower than 5‰, including the peak at 344 keV not used in calibrations and more subject to tailing effect, small fluctuations, and miscalibration problems. In Figure \ref{fig:Residuals values per date}, the relative residuals (Equation \ref{formula relative res}) of one of the detectors through the various calibrations are reported.

\noindent Finally, also the resolutions resulted to be constant in time without any loss of performance. To take into account the strong tailing effect, especially in high energy peaks, the resolution was estimated through the evaluation of the Full Width Half Maximum (FWHM) for every peak. In Table \ref{resolutions}, the values of the FWHM for one of the detectors between different calibrations are reported.

\begin{table}[ht] 
\caption{Resolutions for one of the detectors in the different runs. \vspace{0.5cm} \label{resolutions}}
\begin{tabular}{ccccc}
\toprule
\textbf{Energy (keV)}    & \textbf{FWHM in 05/06-07 (\%)}    & \textbf{FWHM in 05/09-10 (\%)} & \textbf{FWHM in 05/31 (\%)} & \textbf{FWHM in 06/18 (\%)}\\
\midrule
40 keV  & 10.31 $\pm$ 0.01 & 10.01 $\pm$ 0.08 & 9.8 $\pm$ 0.4 & 10.5 $\pm$ 0.1 \\
121 keV & 4.177 $\pm$ 0.007 & 4.11 $\pm$ 0.05 & 4.1 $\pm$ 0.1 & 4.31 $\pm$ 0.05 \\
244 keV & 3.20 $\pm$ 0.03 & 3.15 $\pm$ 0.08 & 3 $\pm$ 1 & 3.4 $\pm$ 0.7 \\
344 keV & 3.19 $\pm$ 0.01 & 2.87 $\pm$ 0.04 & 3.4 $\pm$ 0.3 & 4.31 $\pm$ 0.06 \\
\end{tabular}
\end{table}
%

\noindent From the table, it is clear that the resolution does not get notably worse even in months of data taking, confirming the stability of the whole apparatus in more than a month.

\noindent This study also confirmed that the calibration system used for the CZT detectors is good and guarantees promising performances for the physics run.

\subsection{In-beam Fit}
Finally, the performances of the detectors in a run with the beam on and the radioactive source were studied. The background in DA$\Phi$NE was modeled as:
\begin{equation}
    f_{bkg}(x) = a + b \cdot x + c \cdot \exp(d \cdot x) + \text{erfc}(\frac{e-x}{f}).
\end{equation}
The exponential behavior is due to the fact that the principal background comes from X-rays with an energy lower than 500 keV that lose energy interacting via Compton scattering with the air and the materials between the beam pipe and the detector. The component described by the complementary error function represents a cut on the lower energies brought by the presence of the shielding and the electronics. Beside these two components, a linear function that represents the electronic noise was also added.

The most prominent peaks are the ones coming from the lead scintillation (K$_\alpha$ and K$_\beta$) in the shielding, together with a peak around 200 keV, due to the back-scattering processes described before. The Eu1, Eu2, and Eu3 peaks coming from the source's decays are also prominent over the background. Finally, a peak at 511 keV was observed; this peak comes from the annihilation processes in the surrounding material.

\begin{figure}[ht]
\centering
\includegraphics[width=17cm]{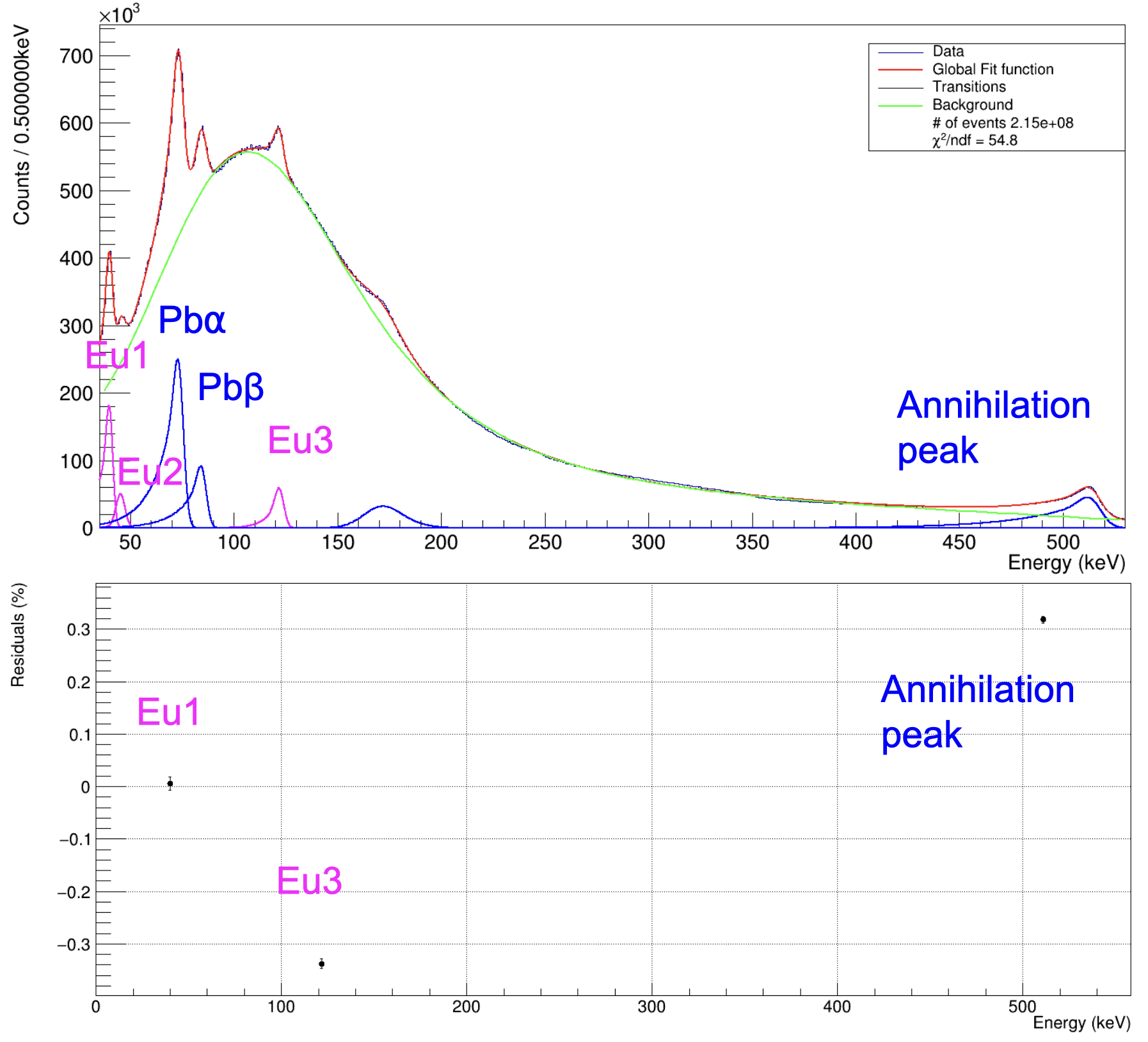}
\caption{Fit of the spectrum acquired with the commercial detectors during the data taking with the source and beam on (up) and relative residuals of the $^{152}$Eu peaks at 40 keV and 121 kev, and for the 511 keV peak.} \label{fig:Fit_old}
\end{figure}

A figure of the final spectrum collected for the detectors made by REDLEN is reported in Figure \ref{fig:Fit_old} up. The spectrum was calibrated using the results of the calibration held between 06th of May 2024 and 07th of May 2024. In Figure \ref{fig:Fit_old} down, the residuals obtained in the fit for the Eu1, Eu3, and annihilation peaks are reported.
Looking at the residuals, they resulted to be around ‰ level in a wide range (from 40 to 511 keV), without a loss in resolutions (11\% at 40 keV, 5\% at 121 keV), confirming the appealing features of the detector in kaonic atoms researches.

The obtained residuals can also be intended as a first estimation of the systematic error that the detectors have during a physics run and are an important indication for the upcoming kaonic atoms measurement.

\section{Discussion}

The studies presented in this article show that the new CZT detection system developed by the SIDDHARTA-2 collaboration shows that this detector is ideal to perform X-ray measurements in a high-rate environment such as the one of the DA$\Phi$NE collider and that it is ready to be used to perform kaonic atom measurements. 

The study on the short-term stability of the detectors showed that the outcome signals of the complex apparatus do not depend on environmental conditions, confirming the perfect stability of these kinds of detectors also after being exposed at high rates.

The study of the long-term stability showed that the system is extremely stable also after switching off the apparatus and the machine and after changing some detectors, confirming the robustness of the electronics and the hardware and software data acquisition. It was also extremely important because it gave a confirmation of the fact that a single long calibration every two of three weeks, even a month, is sufficient to control accurately the systematics and to obtain precise results on the kaonic atoms observables. It was also shown that the high rate due to the environment does not affect at all the detector properties. 

Finally, thanks to the spectrum collected with beam on, it was proved that the good linearity of the detectors, demonstrated in the previous studies, allows to sum the spectra of four different detectors without any loss of performances in terms of resolutions and precision. This run, thanks to the relative residuals collected in the in-beam spectrum, calibrated with the spectra collected with the beam off, was also important to predict a systematic error on the physics measurements under 5‰, a promising result for the foreseen studies on intermediate mass kaonic atoms shift and width, important observables for the understanding of the hadrons low-energy strong interactions with strangeness.
This work was also important to study the ideal fitting function to use for spectroscopic measurement using CZT detectors in a collider, a significant step in the progress of this promising X-ray detection system and in its application in the field of particle and nuclear physics.

This test, together with the past (\cite{abbene_new_2023, scordo_cdznte_2024}) confirm that the new CZT detector developed by the SIDDHARTA-2 collaboration has appealing features to perform kaonic atom measurement for transitions in the range from 20 keV up to 400 keV. This is also the first application of Cadmium-Zinc-Telluride detectors in an accelerator, and these results pave the way for new, important application of this unique semiconductor in the field of particle and nuclear physics.

\vspace{6pt} 

\section*{Acknowledgements}

Part of this work was supported by the Austrian Science Fund (FWF): [P24756-N20 and P33037-N]; the EXOTICA project of the Minstero degli Affari Esteri e della Cooperazione Internazionale, PO22MO03; the Croatian Science Foundation under the project IP-2018–01-8570; the EU STRONG-2020 project (Grant Agreement No. 824093); the EU Horizon 2020 project under the MSCA (Grant Agreement 754496); the Japan Society for the Promotion of Science JSPS KAKENHI Grant No. JP18H05402; the SciMat and qLife Priority Research Areas budget under the program Excellence Initiative - Research University at the Jagiellonian University, and the Polish National Agency for Academic Exchange (Grant No. PPN/BIT/2021/1/00037); the EU Horizon 2020 research and innovation programme under project OPSVIO (Grant Agreement No. 101038099). This work was also supported by the Italian Ministry for University and Research (MUR), under PRIN 2022 PNRR project CUP: B53D23024100001

We thank H. Schneider, L. Stohwasser, and D. Pristauz-Telsnigg from Stefan Meyer-Institut for their fundamental contribution in designing and building the SIDDHARTA-2 setup. We thank as well the INFN, INFN-LNF and the DA$\Phi$NE staff in particular to Dr. Catia Milardi for the excellent working conditions and permanent support.
Catalina Curceanu acknowledge University of Adelaide, where part of this work was done (under the George Southgate fellowship, 2024).



\bibliographystyle{JHEP}
\bibliography{ref}

\providecommand{\href}[2]{#2}\begingroup\raggedright\begin{thebibliography}{10}

\bibitem{del_sordo_progress_2009}
S.~Del~Sordo and {others}, \emph{Progress in the {Development} of {CdTe} and {CdZnTe} {Semiconductor} {Radiation} {Detectors} for {Astrophysical} and {Medical} {Applications}}, \href{https://doi.org/10.3390/s90503491}{\emph{Sensors} {\bfseries 9} (2009) 3491}.

\bibitem{abbene_high-rate_2015}
L.~Abbene and G.~Gerardi, \emph{High-rate dead-time corrections in a general purpose digital pulse processing system}, \href{https://doi.org/10.1107/S1600577515013776}{\emph{Journal of Synchrotron Radiation} {\bfseries 22} (2015) 1190}.

\bibitem{abbene_development_2017}
L.~Abbene, G.~Gerardi, G.~Raso, F.~Principato, N.~Zambelli, G.~Benassi et~al., \emph{Development of new {CdZnTe} detectors for room-temperature high-flux radiation measurements}, \href{https://doi.org/10.1107/S1600577517000194}{\emph{J Synchrotron Rad} {\bfseries 24} 429}.

\bibitem{abbene_room-temperature_2020}
L.~Abbene, F.~Principato, G.~Gerardi, A.~Buttacavoli, D.~Cascio, M.~Bettelli et~al., \emph{Room-temperature x-ray response of cadmium–zinc–telluride pixel detectors grown by the vertical bridgman technique}, .

\bibitem{vicini_optimization_czt_2023}
V.V.Z.S.S.A.N.~{et al.}, \emph{Optimization of quasi-hemispherical cdznte detectors by means of first principles simulation}, {\emph{Sci Rep} {\bfseries 13} (2023) 3212}.

\bibitem{abbene_potentialities_2023}
L.~Abbene, A.~Buttacavoli, F.~Principato, G.~Gerardi, M.~Bettelli, A.~Zappettini et~al., \emph{Potentialities of cdznte quasi-hemispherical detectors for hard x-ray spectroscopy of kaonic atoms at the da$\phi$ne collider}, \href{https://doi.org/10.3390/s23177328}{\emph{Sensors} {\bfseries 23} (2023) }.

\bibitem{abbene_new_2023}
L.~Abbene and {others}, \emph{New opportunities for kaonic atoms measurements from {CdZnTe} detectors}, \href{https://doi.org/10.1140/epjs/s11734-023-00881-x}{\emph{The European Physical Journal Special Topics} {\bfseries 232} (2023) 1487}.

\bibitem{scordo_cdznte_2024}
A.~Scordo and {others}, \emph{{CdZnTe} detectors tested at the {DA$\Phi$NE} collider for future kaonic atoms measurements}, \href{https://doi.org/10.1016/j.nima.2023.169060}{\emph{Nuclear Instruments and Methods in Physics Research Section A: Accelerators, Spectrometers, Detectors and Associated Equipment} {\bfseries 1060} (2024) 169060}.

\bibitem{bazzi_first_2011}
M.~Bazzi and {others}, \emph{First measurement of kaonic helium-3 {X}-rays}, \href{https://doi.org/10.1016/j.physletb.2011.02.001}{\emph{Physics Letters B} {\bfseries 697} (2011) 199}.

\bibitem{bazzi_new_2011}
M.~Bazzi and {others}, \emph{A new measurement of kaonic hydrogen {X}-rays}, \href{https://doi.org/10.1016/j.physletb.2011.09.011}{\emph{Physics Letters B} {\bfseries 704} (2011) 113}.

\bibitem{curceanu_kaonic_2020}
C.~Curceanu and {others}, \emph{Kaonic {Deuterium} {Measurement} with {SIDDHARTA}-2 on {DA}${\Phi} ${NE}}, \href{https://doi.org/10.5506/APhysPolB.51.251}{\emph{Acta Physica Polonica B} {\bfseries 51} (2020) 251}.

\bibitem{sirghi_kaonic_helium_2022}
D.~Sirghi, F.~Sirghi, F.~Sgaramella, M.~Bazzi, D.~Bosnar, M.~Bragadireanu et~al., \emph{A new kaonic helium measurement in gas by siddhartino at the da$\phi$ne collider*}, \href{https://doi.org/10.1088/1361-6471/ac5dac}{\emph{Journal of Physics G: Nuclear and Particle Physics} {\bfseries 49} (2022) 055106}.

\bibitem{curceanu_frontiers_2023}
C.~Curceanu and {others}, \emph{Kaonic atoms at the da$\phi$ne collider: a strangeness adventure}, \href{https://doi.org/10.3389/fphy.2023.1240250}{\emph{Frontiers in Physics} {\bfseries 11} (2023) }.

\bibitem{curceanu_modern_2019}
C.~Curceanu and {others}, \emph{The modern era of light kaonic atom experiments}, \href{https://doi.org/10.1103/RevModPhys.91.025006}{\emph{Reviews of Modern Physics} {\bfseries 91} (2019) 025006}.

\bibitem{artibani_odyssey_2024}
F.~Artibani and {others}, \emph{The odyssey of kaonic atoms studies at the da{$\Phi$}ne collider: From dear to siddharta-2}, {\emph{Acta Physica Polonica B} {\bfseries 55} (2024) 5}.

\bibitem{bernard_chiral_2008}
V.~Bernard, \emph{Chiral {Perturbation} {Theory} and {Baryon} {Properties}}, \href{https://doi.org/10.1016/j.ppnp.2007.07.001}{\emph{Progress in Particle and Nuclear Physics} {\bfseries 60} (2008) 82}.

\bibitem{cieply_pole_2016}
A.~Cieplý, M.~Mai, U.-G.~Meißner and J.~Smejkal, \emph{On the pole content of coupled channels chiral approaches used for the {K}¯{N} system}, \href{https://doi.org/10.1016/j.nuclphysa.2016.04.031}{\emph{Nuclear Physics A} {\bfseries 954} (2016) 17}.

\bibitem{torigoe_atomic_cascade_1977}
E.~Torigoe, \emph{The atomic cascade in kaonic atoms}, \href{https://doi.org/https://doi.org/10.1016/0003-4916(77)90225-1}{\emph{Annals of Physics} {\bfseries 105} (1977) 1}.

\bibitem{ramos_kaon_condensation_neutron_stars_2000}
A.~Ramos, J.~Schaffner-Bielich and J.~Wambach, \emph{Kaon condensation in neutron stars}, .

\bibitem{Bosnar_kaonic_lead_2024}
D.~Bosnar et~al., \emph{{Kaonic lead feasibility measurement at DA\ensuremath{\Phi}NE to solve the charged kaon mass discrepancy}},  \href{https://arxiv.org/abs/2405.12942}{{\ttfamily 2405.12942}}.

\bibitem{friedman_density-dependent_1994}
E.~Friedman, A.~Gal and C.J.~Batty, \emph{Density-dependent {K}- nuclear optical potentials from kaonic atoms}, \href{https://doi.org/10.1016/0375-9474(94)90921-0}{\emph{Nuclear Physics A} {\bfseries 579} (1994) 518}.

\bibitem{batty_strong_1997}
C.~Batty, E.~Friedmann and A.~Gal, \emph{Strong interaction physics from hadronic atoms}, .

\bibitem{bazzi_kaonic_helium_2009}
M.~Bazzi, G.~Beer, L.~Bombelli, A.~Bragadireanu, M.~Cargnelli, G.~Corradi et~al., \emph{Kaonic helium-4 x-ray measurement in siddharta}, \href{https://doi.org/https://doi.org/10.1016/j.physletb.2009.10.052}{\emph{Physics Letters B} {\bfseries 681} (2009) 310}.

\bibitem{tolos_strangeness_2020}
L.~Tolos and L.~Fabbietti, \emph{Strangeness in nuclei and neutron stars}, \href{https://doi.org/10.1016/j.ppnp.2020.103770}{\emph{Progress in Particle and Nuclear Physics} {\bfseries 112} (2020) 103770}.

\bibitem{milardi_present_2009}
C.~Milardi and {others}, \emph{Present status of the {DAFNE} upgrade and perspectives}, \href{https://doi.org/10.1142/S0217751X09043687}{\emph{Int. J. Mod. Phys. A} {\bfseries 24} (2009) 360}.

\bibitem{Milardi:2021khj}
C.~Milardi et~al., \emph{{DA\ensuremath{\Phi}NE Commissioning for SIDDHARTA-2 Experiment}}, \href{https://doi.org/10.18429/JACoW-IPAC2021-TUPAB001}{\emph{JACoW} {\bfseries IPAC2021} (2021) TUPAB001}.

\bibitem{Milardi:2024efr}
C.~Milardi et~al., \emph{{DAFNE operation strategy for the observation of the kaonic deuterium}}, \href{https://doi.org/10.18429/JACoW-IPAC2024-WEPR17}{\emph{JACoW} {\bfseries IPAC2024} (2024) WEPR17}.

\bibitem{boscolo_touschek_2007}
M.~Boscolo, M.~Biagini, S.~Guiducci and P.~Raimondi, \emph{Touschek background and beam lifetime studies for the {DAFNE} upgrade},  in \emph{2007 {IEEE} Particle Accelerator Conference ({PAC})}, pp.~1454--1456, \href{https://doi.org/10.1109/PAC.2007.4440787}{DOI}.

\bibitem{Abbene_energy_2013}
L.~Abbene, G.~Gerardi, G.~Raso, S.~Basile, M.~Brai and F.~Principato, \emph{Energy resolution and throughput of a new real time digital pulse processing system for x-ray and gamma ray semiconductor detectors}, \href{https://doi.org/10.1088/1748-0221/8/07/P07019}{\emph{Journal of Instrumentation} {\bfseries 8} (2013) P07019}.

\bibitem{Gerardi_digital_2014}
G.~Gerardi and L.~Abbene, \emph{A digital approach for real time high-rate high-resolution radiation measurements}, \href{https://doi.org/https://doi.org/10.1016/j.nima.2014.09.047}{\emph{Nuclear Instruments and Methods in Physics Research Section A: Accelerators, Spectrometers, Detectors and Associated Equipment} {\bfseries 768} (2014) 46}.

\bibitem{skurzok_characterization_2020}
M.~Skurzok, A.~Scordo, S.~Niedzwiecki, A.~Baniahmad, M.~Bazzi, D.~Bosnar et~al., \emph{Characterization of the {SIDDHARTA}-2 luminosity monitor}, .

\bibitem{gysel_implementation_spectrum_fitting_2003}
M.~Gysel, P.~Lemberge and P.~Van~Espen, \emph{Implementation of a spectrum fitting procedure using a robust peak model}, \href{https://doi.org/10.1002/xrs.666}{\emph{X‐Ray Spectrometry} {\bfseries 32} (2003) 434 }.

\bibitem{borrella_peak_shape_2021}
A.~Borella, M.~Bruggeman, R.~Rossa and P.~Schillebeeckx, \emph{Peak shape calibration of a cadmium zinc telluride detector and its application for the determination of uranium enrichment}, \href{https://doi.org/https://doi.org/10.1016/j.nima.2020.164718}{\emph{Nuclear Instruments and Methods in Physics Research Section A: Accelerators, Spectrometers, Detectors and Associated Equipment} {\bfseries 986} (2021) 164718}.

\bibitem{redus_characterization_cdte_2009}
R.~Redus, J.~Pantazis, T.~Pantazis, A.~Huber and B.~Cross, \emph{Characterization of cdte detectors for quantitative x-ray spectroscopy}, \href{https://doi.org/10.1109/TNS.2009.2024149}{\emph{Nuclear Science, IEEE Transactions on} {\bfseries 56} (2009) 2524 }.

\end{thebibliography}\endgroup

\end{document}